\documentclass[twocolumn,english,aps, pra, eqsecnum]{revtex4-1}
\usepackage[T1]{fontenc}
\usepackage[latin9]{inputenc}
\usepackage{bm}
\usepackage{amsmath}
\usepackage{amssymb}
\usepackage{graphicx}
\usepackage{babel}
\begin{document}

\title{Coherent states in projected Hilbert spaces}

\author{P. D. Drummond and M. D. Reid}

\affiliation{Centre for Quantum and Optical Science, Swinburne University of Technology,
Melbourne 3122 Australia}
\begin{abstract}
Coherent states in a projected Hilbert space have many useful properties.
When there are conserved quantities, a representation of the entire
Hilbert space is not necessary. The same issue arises when conditional
observations are made with post-selected measurement results. In these
cases, only a part of the Hilbert space needs to be represented, and
one can define this restriction by way of a projection operator. Here
coherent state bases and normally-ordered phase-space representations
are introduced for treating such projected Hilbert spaces, including
existence theorems and dynamical equations. These techniques are very
useful in studying novel optical or microwave integrated photonic
quantum technologies, such as boson sampling or Josephson quantum
computers. In these cases states become strongly restricted due to
inputs, nonlinearities or conditional measurements. This paper focuses
on coherent phase states, which have especially simple properties.
Practical applications are reported on calculating recurrences in
anharmonic oscillators, the effects of arbitrary phase-noise on Schr\"odinger
cat fringe visibility, and on boson sampling interferometry for large
numbers of modes.
\end{abstract}
\maketitle

\section{Introduction}

Coherent states are a widely used concept in quantum physics. Originally
introduced by Schr\"odinger\cite{Schrodinger_CS}, and extended by
Bargmann\cite{Bargmann:1961} and Glauber\cite{Glauber1963-states},
these are often applied to quantum optics and quantum technologies.
The generalized P-representation \cite{Chaturvedi:1977,Drummond_Gardiner_PositivePRep}
makes extensive use of these states: it is a complete phase-space
representation on multi-mode bosonic Hilbert space, which extends
the Glauber-Sudarshan P-function \cite{Glauber_1963_P-Rep,Sudarshan_1963_P-Rep}
by allowing non-singular distributions for any quantum state or density
matrix. It can be utilized for both exact, analytic solutions, and
stochastic simulations of dynamics\cite{Hillery_Review_1984_DistributionFunctions,Gardiner_Book_QNoise,drummond2016quantum}.
It has a high degree of scalability, permitting exponentially large
Hilbert space problems to be treated \cite{Deuar:2007_BECCollisions},
as well as large occupation numbers \cite{Dowling2005}. 

Yet sometimes only part of the full Hilbert space is needed. There
may only be occupation numbers of $\left(0,1\right)$, or else conservation
laws of energy, number or momentum may restrict the Hilbert space.
For these physical problems, it is unnecessary to utilize states that
form a basis for the whole Hilbert space, since only part of it is
occupied. This type of problem can be treated using projected representations.
An example of this is number conservation, which can lead to invariance
under projection such that states with total numbers differing from
$N$  do not need to be represented. 

To treat such restricted Hilbert spaces in bosonic cases, this paper
introduces projected coherent states and corresponding phase-space
representations where projections are included by redefining the coherent
states. The resulting mappings are more focused and compact than with
the full coherent basis. This approach includes a type of discrete
coherent state, called a coherent phase states (CPS), which are cousins
to the phase states of quantum optics \cite{barnett1989hermitian,Barnett:1993}.
These allow a definition of a discrete P-function with similarities
to the discrete Wigner function \cite{wootters1987wigner,gibbons2004discrete,chaturvedi2010wigner}.
The utility of coherent phase states is that they satisfy operator
identities almost identical to the full coherent states, thus allowing
dynamical evolution to be calculated. They can be applied efficiently
to photonic networks and boson sampling problems \cite{Opanchuk2016-quantum}.

A projected phase-space technique can have a greater numerical efficiency,
and lower sampling error, when the physically occupied Hilbert space
is a small fraction of the Hilbert space. As an example of this, fermionic
phase-space representations of the Hubbard model that include conservation
laws result in improved numerical simulations \cite{Imada:2007_2DHM_Superconductivity}.
In other cases where the total experimental density matrix $\hat{\rho}$
is not invariant under projection, one may still be interested in
measurements only on a projected part of the density matrix. These
are post-selected measurements, and are common in many quantum optics
experiments \cite{Aspect:1981_PRL47,ReidqubitPhysRevA.90.012111,rosales2014probabilistic}. 

The examples given here are all cases where the projectors commute
with the number operator. A typical experimental protocol is the photonic
Bell experiment, where all measurements yielding the vacuum state
are projected out via post-selection \cite{ZeilingerBellPhysRevLett.81.5039,rosales2014probabilistic}.
There are many other novel experimental technologies which combine
growing exponential complexity with a restricted Hilbert space. A
well-known case is ``boson sampling''~\cite{AaronsonArkhipov:2011,AaronsonArkhipov2013LV},
which has led to photonic waveguide experiments \cite{Broome2013,Crespi2013,Tillmann2013,Spring72013}
and metrology proposals~\cite{OBrienMetrology,Motes2015_PRL114}.
Some of these ideas, originally in the optical domain, have also been
extended to superconducting waveguide qubits \cite{CorcolesIBM2015},
and are applicable to massive bosons as well. 

Other quantum technologies, including arrays of optomechanical devices
\cite{eichenfield2009picogram} and quantum gas microscopes for BEC
systems \cite{bakr2009quantum}, can operate in a similar regime.
These systems have a combination of exponential complexity - making
orthogonal basis techniques difficult to scale - together with restricted
occupation numbers which makes projection techniques useful. This
paper establishes the general foundations of this approach. Particular
applications are treated elsewhere \cite{Opanchuk2016-quantum}.

Section II introduces projected coherent states, together with coherent
phase states. Differential and matrix identities are obtained in Section
III. Section IV defines projected P-representations, which use the
projected coherent states, and obtains existence theorems while Section
V discusses applications to photonic networks, including an anharmonic
example, phase decoherence in a Schr\"odinger Cat, and boson sampling
experiments. Finally, Section VI gives the conclusions.

\section{projected states}

Coherent states have many uses in calculations in physics, and they
are a complete basis. However, dealing with the complete Hilbert space
is often not necessary. There can be restrictions and symmetries that
limit the available states. Here, ways to define coherent states with
projective restrictions are introduced. These still lead to a complete
basis in the Hilbert sub-space of interest. 

Projection operators on a Hilbert space can be applied to states,
operators or to the density matrix. Suppose the projector is $\mathcal{P}$,
defined so that $\mathcal{P}=\mathcal{P}^{2}$. If there are multiple
projections $\mathcal{P}_{j}$, they are assumed to commute, so that
their product $\mathcal{P}=\prod\mathcal{P}_{j}$ is a projector.
Given a state $\left|\psi\right\rangle $, its projected versions
will be written as $\left|\psi\right\rangle _{\mathcal{P}}=\mathcal{P}\left|\psi\right\rangle $.
When there are normalization requirements, there is an additional
scale factor, defined as needed. In this section, results are obtained
for projection operators applied to coherent states.

\subsection{Coherent states}

Throughout this paper, an $M$  mode system of bosons is treated with
number states that are outer products of number states $\left|n_{j}\right\rangle _{j}$
for each mode, so that: 
\begin{equation}
\left|n_{1},\ldots,n_{M}\right\rangle \equiv\prod_{j=1}^{M}\left|n_{j}\right\rangle _{j}\,.
\end{equation}

The coherent states $\left\Vert \bm{\alpha}\right\rangle $ and $\left|\bm{\alpha}\right\rangle $
are respectively the unnormalized and normalized eigenstate of annihilation
operators $\hat{\mathbf{a}}=\left[\hat{a}_{1},\ldots\hat{a}_{M}\right]$,
with eigenvalues $\bm{\alpha}$. To prevent ambiguity, coherent states
are labelled with greek letters $\left|\bm{\alpha}\right\rangle $,
number states with roman letters $\left|\bm{n}\right\rangle $. The
coherent state basis is conventionally written in terms of $\bm{\alpha}=\left[\alpha_{1},\ldots\alpha_{M}\right]$,
which is a complex vector of coherent amplitudes. In the multi-mode
case, the unnormalized coherent state has the standard form of \cite{Bargmann:1961}:
\begin{eqnarray}
\left\Vert \bm{\alpha}\right\rangle  & = & \sum_{\mbox{n}\ge0}\prod_{i}\frac{\alpha_{i}^{n}}{\sqrt{n_{i}!}}\left|n_{i}\right\rangle _{i}=\sum_{\mbox{n}\ge0}\left\Vert \bm{\alpha}\right\rangle _{\mathbf{n}}.
\end{eqnarray}
where the state $\left\Vert \bm{\alpha}\right\rangle _{\mathbf{n}}$
is an unnormalized coherent state projected onto a fixed number $\bm{n}$,
so that:
\begin{equation}
\left\Vert \bm{\alpha}\right\rangle _{\mathbf{n}}\equiv\prod_{i}\frac{\alpha_{i}^{n}}{\sqrt{n_{i}!}}\left|n_{i}\right\rangle _{i}\,.
\end{equation}

For coherent states projected onto single number states, the following
identities hold:
\begin{align}
\hat{a}\left\Vert \alpha\right\rangle _{n} & =\alpha\left\Vert \alpha\right\rangle _{n-1}\,,\label{eq:coherent-projection-identity}\\
\hat{a}^{\dagger}\left\Vert \alpha\right\rangle _{n} & =\frac{n+1}{\alpha}\left\Vert \alpha\right\rangle _{n+1}.\nonumber 
\end{align}
The corresponding \emph{normalized }coherent state is \cite{glauber1963photon,Glauber1963_CoherentStates}:
\begin{equation}
\left|\bm{\alpha}\right\rangle =g\left(\bm{\alpha}\right)^{-1}\left\Vert \bm{\alpha}\right\rangle \,,\label{eq:normalized-coherent-state}
\end{equation}
where $g\left(\bm{\alpha}\right)=e^{\left|\bm{\alpha}\right|^{2}/2}=\sqrt{\left\langle \bm{\alpha}\right.\left\Vert \bm{\alpha}\right\rangle }$
is the norm.

\subsection{Projected coherent states}

Next, consider a \emph{projected} coherent basis. In the unnormalized
case this is simply
\begin{equation}
\left\Vert \bm{\alpha}\right\rangle _{\mathcal{P}}=\mathcal{P}\left\Vert \bm{\alpha}\right\rangle \,.
\end{equation}
 For the normalized case, there is a modified normalization as well,
so that in this case one can define:
\begin{eqnarray}
\left|\bm{\alpha}\right\rangle _{\mathcal{P}} & = & g_{\mathcal{P}}^{-1}\left\Vert \bm{\alpha}\right\rangle _{\mathcal{P}}\,,\label{eq:projected-norm-coherent}
\end{eqnarray}
 where $g_{\mathcal{P}}=g_{\mathcal{P}}\left(\bm{\alpha}\right)=\sqrt{\left\langle \bm{\alpha}\right\Vert \mathcal{P}\left\Vert \bm{\alpha}\right\rangle }$.
The abbreviated form $g_{\mathcal{P}}$ for the normalization will
be used if there is no ambiguity. This is not just an outer product,
but rather is a linear combination of all possible number states $\left|\bm{n}\right\rangle $
that satisfy the projection requirements, with coefficients given
by the appropriate coherent amplitudes. 

\begin{figure}
\centering{}\includegraphics[clip,width=0.9\columnwidth]{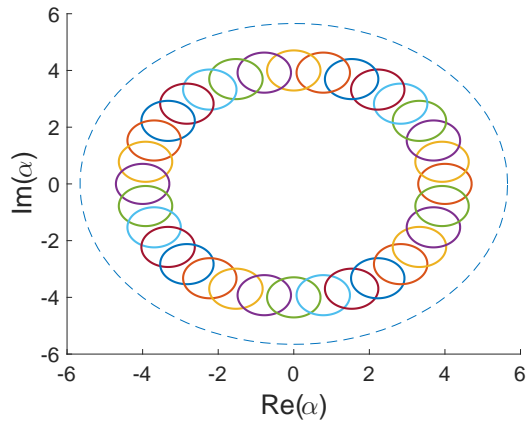}\caption{The coherent phase states for dimension $d=32,$ with coherent radius
of $\alpha=4$. The dotted line is the number state cut-off at a radius
of $\sqrt{d}$ . The colored circles are a circle of uncertainty for
each coherent phase-state of radius $\Delta x=\Delta y=1/\sqrt{2}$.
This is the Heisenberg uncertainty or standard deviation in a measurement
of $\hat{x}=\left(\hat{a}+\hat{a}^{\dagger}\right)/2$ for a coherent
state $\left|\alpha\right\rangle =\left|x+iy\right\rangle $.\label{figCoherent phase state}}
\end{figure}

As one example of this, coherent phase states (CPS) will be introduced.
These are a complete, linearly independent set of projected coherent
states which are complementary to a set of number states, as illustrated
in Fig (\ref{figCoherent phase state}). In general, projected states
can have both coherence properties and a conservation law. Such properties
are useful in describing Bose Einstein condensates which have long
range coherence, but a fixed or bounded number. 

\subsection{Number projected coherent states}

Many common physical systems have a restricted range of bosonic occupation
numbers $n_{i}$, such that $\bm{n}\in S$, where $S$ is the set
of physically allowed values. As one example, the total boson number
may be constrained so that $\sum n_{i}=N$. This can occur due to
number conservation combined with a number-sensitive preparation or
post-selection. If there are dissipative losses, the total number
may have an upper bound so that $\sum n_{i}\leq N_{m}$. 

To treat this, define $\mathcal{P}\equiv\mathcal{N}$ as the projector
onto the number states with $\bm{n}\in S$. This is an identity operator
in the projected space, so that: 
\begin{equation}
\hat{I}_{\mathcal{N}}=\mathcal{N}=\sum_{\bm{n}\in S}\left|\bm{n}\right\rangle \left\langle \bm{n}\right|.
\end{equation}
The unnormalized projected state is :
\begin{equation}
\left\Vert \bm{\alpha}\right\rangle _{\mathcal{\mathcal{N}}}=\mathcal{N}\left\Vert \bm{\alpha}\right\rangle =\sum_{\bm{n}\in S}\prod_{i}\frac{\alpha_{i}^{n_{i}}}{\sqrt{n_{i}!}}\left|n_{1},\ldots,n_{M}\right\rangle \,.\label{eq:number-projected-unnorm-coherent}
\end{equation}

The normalized projected coherent state is then defined as in (\ref{eq:projected-norm-coherent}),
where the normalization is:
\begin{equation}
g_{\mathcal{\mathcal{N}}}\left(\bm{\alpha}\right)=\sqrt{\sum_{\bm{n}\in S}\prod_{i}\frac{\left|\alpha_{i}\right|^{2n_{i}}}{n_{i}!}}\,.
\end{equation}
For the single-mode case with a number cutoff projection such that
$n\leq N_{m}=n_{m}$, the normalization is proportional to an upper
incomplete gamma function \cite{zwillinger2014table},
\begin{equation}
g_{\mathcal{N}}\left(\alpha\right)=\sqrt{\frac{1}{n_{m}!}e^{|\alpha|^{2}}\Gamma\left(1+n_{m},|\alpha|^{2}\right)}\,.\label{eq:gamma-function-norm}
\end{equation}

In cases of number projections, invariant observables of form $\hat{O}=\hat{a}_{m}^{\dagger}\cdots\hat{a}_{n}$
are usually of most interest. These have an equal number of annihilation
and creation operators, and are called \emph{number-conserving observables}. 

When both the projectors and observables are number-conserving, it
follows that the observables commute with the projectors, as they
have a complete set of simultaneous eigenstates, i.e., for number
conserving observables,
\begin{equation}
\hat{a}_{m}^{\dagger}\cdots\hat{a}_{n}\mathcal{N}=\mathcal{N}\hat{a}_{m}^{\dagger}\cdots\hat{a}_{n}\,.
\end{equation}
In some, but not all cases, invariant observables have the number
states as eigenstates, so that $\hat{O}\left|\mathbf{n}\right\rangle =O_{\mathbf{n}}\left|\mathbf{n}\right\rangle .$

\subsection{Qudit bosonic coherent states\label{subsec:Qudit-bosonic-coherent}}

A qudit coherent state is defined to have a range of occupation numbers
$n_{i}$ in \emph{each} mode, such that $n_{i}\in s$, where $s$
is a set of $d$ allowed number values. This type of finite dimensional
qudit \cite{Dennison2001PRA} has a projector denoted $\mathcal{Q}$. 

The projected identity operator factorizes, so that:
\begin{equation}
\hat{I}_{\mathcal{Q}}=\mathcal{Q}=\prod_{j}\left[\sum_{n_{j}\in s}\left|n_{j}\right\rangle \left\langle n_{j}\right|_{j}\right]\,.\label{eq:projected-qudit-identity}
\end{equation}

An alternative approach in this situation would be to define $SU(d)$
coherent states for qudits. These have different combinatorial coefficients
\cite{Perelomov_1972_Coherent_states_LieG,Perelomov_book_Coherent_state,Review_coherent_states_Gilmore_1990,Arvind:1998,Barry_PD_qubit_SU}.
However, this is a Lie group symmetry which is more suitable for spins
rather than for boson operator identities, due to the different operator
algebras in the two cases. 

Binary qubits occur for for $d=2$, where the projected coherent states
are  isomorphic to $SU(2)$ coherent states. The simplest case has
$n=0,1$; so that only zero and one boson states are included, which
makes this useful for Bell inequalities, quantum computers or boson
sampling. 

For general qudits, the single mode case is identical to the previous
section. In the multi-mode case, the unnormalized state factorizes
so that:
\begin{equation}
\left\Vert \bm{\alpha}\right\rangle _{\mathcal{Q}}=\mathcal{Q}\left\Vert \bm{\alpha}\right\rangle =\prod_{i}\left[\sum_{n_{i}\in s}\frac{\alpha_{i}^{n_{i}}}{\sqrt{n_{i}!}}\left|n_{i}\right\rangle _{i}\right]\,.\label{eq: finite-dimensional}
\end{equation}
 The normalized multi-mode qudit coherent state is given in (\ref{eq:projected-norm-coherent}),
with normalization $g_{\mathcal{Q}}\left(\bm{\alpha}\right)=\prod_{i}g_{\mathcal{Q}}\left(\alpha_{i}\right)$,
where
\begin{equation}
g_{\mathcal{Q}}\left(\alpha\right)=\sqrt{\sum_{n\in s}\frac{\left(\alpha^{*}\alpha\right)^{n}}{n!}}\,.
\end{equation}
Here $g_{\mathcal{Q}}\left(\alpha\right)$ is given analytically by
(\ref{eq:gamma-function-norm}) in the number cut-off case with $n_{i}\leq n_{m}$.
The approximation $g_{\mathcal{Q}}\left(\bm{\alpha}\right)\approx e^{\left|\bm{\alpha}\right|^{2}/2}$
is valid for number cut-off qudit projections such that $\left|\alpha_{i}\right|^{2}\ll d$.

The single-mode qudit projection operator can be rewritten in terms
of coherent number states as:
\begin{equation}
\hat{I}_{\mathcal{Q}}=\sum_{n\in s}g_{n}^{-2}\left\Vert \alpha\right\rangle _{n}\left\langle \text{\ensuremath{\alpha}}\right\Vert _{n}\,,\label{eq:identity-scaled-number}
\end{equation}
where the normalization coefficient of a coherent state projected
onto a single number state is:
\begin{equation}
g_{n}=\sqrt{\left|\alpha\right|^{2n}/n!}.\label{eq:qudit-scaling-C}
\end{equation}

Here $g_{n}^{2}$ has a well-known role in probability theory: combined
with an overall normalization of $e^{-\left|\alpha\right|^{2}}$,
it is the probability of observing a number $n$ in a Poisson distribution
with mean $\left|\alpha\right|^{2}$. For large $\left|\alpha\right|^{2}$
this is approximately normal, with equal mean and variance, $\mu=\sigma^{2}=\left|\alpha\right|^{2}$.
Hence, in the large $\left|\alpha\right|$ limit, 
\begin{equation}
g_{n}^{2}\approx\frac{1}{\sqrt{2\pi\left|\alpha\right|^{2}}}e^{\left|\alpha\right|^{2}-(n/\left|\alpha\right|-\left|\alpha\right|)^{2}/2}.
\end{equation}

This Gaussian cutoff for $n/\left|\alpha\right|^{2}>1$ is the main
reason why even a projected coherent state can still maintain many
of the useful properties of the full set. This is rapidly convergent,
so in a numerical calculation in which an initial coherent state is
projected, the calculations can be easily repeated with a larger cutoff
to check that no significant errors are introduced.

\subsection{Coherent phase states\label{subsec:Coherent-phase-states}}

Qudit number projections over a contiguous interval $n_{0}\leq n_{i}\leq n_{m}$
have many interesting properties. It is useful to consider coherent
state amplitudes with discrete phases, so $\alpha^{(q)}\equiv\alpha\exp(iq\phi)$,
for $q=0,\ldots d-1$, where 
\begin{equation}
\phi\equiv\frac{2\pi}{d}\,.\label{eq:phase-choice}
\end{equation}
Here $\alpha^{(q)}$ is defined relative to a reference amplitude
$\alpha$. This gives a projected basis with coherent amplitudes having
a fixed intensity, distributed in a circle on the complex plane as
shown in Fig (\ref{figCoherent phase state}), which can be termed
\emph{coherent phase states} (CPS). These types of projected coherent
states are similar to the quantum phase states \cite{barnett1989hermitian,Barnett:1993},
which have a well-defined phase. The simplest has the ground state
included so that $n_{0}=0$.

They have the useful property that discrete Fourier transform relations
are available for a coherent qudit with $d=1+n_{m}-n_{0}$. A circular
basis results in an invertible mapping between coherent phase and
number states for $n_{0}\leq n\leq n_{m}$ , since it can be written
in the form:
\begin{align}
\left\Vert \alpha^{(q)}\right\rangle _{\mathcal{Q}} & =\sum_{n=n_{0}}^{n_{m}}e^{iqn\phi}\left\Vert \alpha\right\rangle _{n}\,\label{eq:inverse-DFT-phase-coherent}
\end{align}
which, from the discrete Fourier transform theorem, has the corresponding
inverse relation that: 
\begin{equation}
\left\Vert \alpha\right\rangle _{n}=\frac{1}{d}\sum_{q=0}^{d-1}e^{-iqn\phi}\left\Vert \alpha^{(q)}\right\rangle _{\mathcal{Q}}\,.
\end{equation}
There is a corresponding multimode coherent phase state, written as:
\begin{equation}
\left\Vert \bm{\alpha}^{(\bm{q})}\right\rangle _{\mathcal{Q}}=\prod_{j=1}^{M}\left\Vert \alpha^{(q_{j})}\right\rangle _{\mathcal{Q}}\,,
\end{equation}
with a normalized form 
\begin{equation}
\left|\bm{\alpha}^{(q)}\right\rangle _{\mathcal{Q}}=g_{\mathcal{Q}}^{-1}\left\Vert \bm{\alpha}^{(\bm{q})}\right\rangle _{\mathcal{Q}}.
\end{equation}
This basis is familiar with qubits having $d=2$ and $n_{0}=0$, where
$g_{\mathcal{Q}}=\sqrt{1+\left|\alpha\right|^{2}}$ , and one has:
\begin{align}
\left|\alpha^{(0,1)}\right\rangle _{\mathcal{Q}} & =g_{\mathcal{Q}}^{-1}\left[\left|0\right\rangle \pm\alpha\left|1\right\rangle \right]\,.
\end{align}

\begin{figure}
\includegraphics[width=0.9\columnwidth]{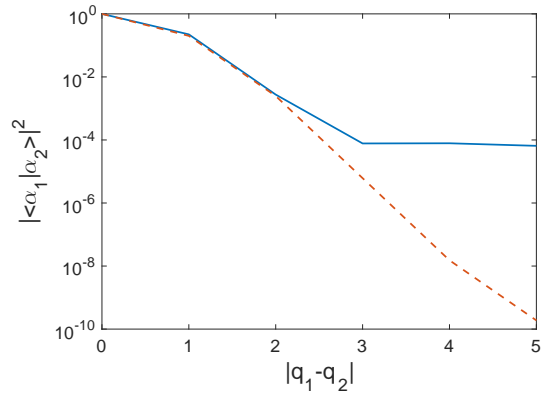}\caption{Orthogonality versus phase difference of coherent phase states $\left|\alpha_{i}\right\rangle =\left|\alpha^{(q_{i})}\right\rangle $
for $n_{0}=0$ and $n_{m}=11$, with $\left|\alpha\right|^{2}=d/2=6$.
Solid line gives the degree of orthogonality of two coherent phase
states with distance $|q_{1}-q_{2}|$, dotted line the orthogonality
of two standard coherent states.\label{fig:Orthogonality-of-coherent}}
\end{figure}

The coherent phase states are linearly independent, but not completely
orthogonal. In general,
\begin{equation}
\left\langle \alpha^{(q_{1})}\right.\left|\alpha^{(q_{2})}\right\rangle _{\mathcal{Q}}=M_{q_{1},q_{2}}\,,
\end{equation}
where the inner product of two coherent phase states is given by:
\begin{equation}
M_{q_{1},q_{2}}=g_{\mathcal{Q}}^{-2}\sum_{n=n_{0}}^{n_{m}}e^{i(q_{2}-q_{1})n\phi}\frac{\left|\alpha\right|^{2n}}{n!}\,.
\end{equation}
With the choice that $\left|\alpha\right|^{2}=(n_{0}+d/2)$, states
with $q_{1}\neq q_{2}$ are orthogonal for the qubit case, and approximately
orthogonal in the general case, as shown in Fig (\ref{fig:Orthogonality-of-coherent}).

\begin{figure}
\includegraphics[width=0.9\columnwidth]{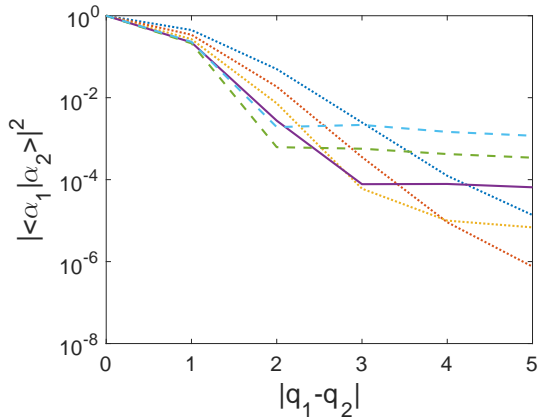}\caption{Orthogonality versus phase difference of coherent phase states $\left|\alpha_{i}\right\rangle =\left|\alpha^{(q_{i})}\right\rangle $
for $n_{0}=0$ and $n_{m}=11$, with $\left|\alpha\right|^{2}=3,4,5,6,7,8$.
The dotted lines have smaller radii with $\left|\alpha\right|^{2}<6$,
with $\left|\alpha\right|^{2}=3$ the upper line. The solid line has
$\left|\alpha\right|^{2}=6$. The dashed lines have larger radii,
with $\left|\alpha\right|^{2}=8$ the upper line on the right.\label{fig:Orthogonality-of-coherent-2}}
\end{figure}

 Such states form a middle ground between the number states, which
are orthogonal but lack coherence, and the coherent states which have
coherence but are highly over-complete. Changing the coherent radius
$r=\left|\alpha\right|$ has no effect on the completeness of the
mapping, but it does change the orthogonality, as shown in Fig (\ref{fig:Orthogonality-of-coherent-2}). 

A smaller radius gives closely spaced coherent amplitudes with reduced
orthogonality for nearest neighbors, but this increases for widely
separated phases. For a larger coherent radius the orthogonality at
small phase separations is greater. However, for large radius \emph{and}
large phase separations, the orthogonality is reduced since the number
cutoff has a strong effect in this limit.

\subsection{Coherent phase expansions\label{subsec:Coherent-phase-states-1}}

Since coherent phase states comprise a complete basis in the projected
subspace, it is possible to expand an arbitrary projected state $\left|\psi\right\rangle =\mathcal{P}\left|\psi\right\rangle $
using these states, just as with the full coherent basis \cite{Glauber_1963_P-Rep}.
This expansion can be written in a multimode case as: 
\begin{equation}
\left|\psi\right\rangle =\sum_{\bm{q}}\psi_{\bm{q}}\left\Vert \bm{\alpha}^{(\bm{q})}\right\rangle _{\mathcal{Q}}\,.\label{eq:CPS-wavefunction-expansion}
\end{equation}
Unlike the non-unique coherent state case, this expansion is unique,
owing to the linear independence of the CPS basis. In the single-mode
case, given a number state expansion $\left|\psi\right\rangle =\sum_{\bm{n}}\psi_{\bm{n}}\left|\bm{n}\right\rangle $,
the CPS coefficients are:
\begin{equation}
\psi_{\bm{q}}=\sum_{\bm{n}}\psi_{\bm{n}}\prod_{j}\frac{\sqrt{n_{j}!}}{d\left(\alpha^{(q_{j})}\right)^{n_{j}}}\,.
\end{equation}
The expectation value of any observable $\hat{O}$ in a pure state
is then given by:
\begin{equation}
\left\langle \hat{O}\right\rangle =\sum_{\overrightarrow{\bm{q}}}\psi_{\bm{q}}^{*}\psi_{\bm{q}'}\left\langle \bm{\alpha}^{(\bm{q})}\right\Vert \hat{O}\left\Vert \bm{\alpha}^{(\bm{q}')}\right\rangle _{\mathcal{Q}}\,.
\end{equation}

As a simple example of this type of expansion, an arbitrary projected
coherent state $\left|\tilde{\alpha}\right\rangle $can be expanded
using a CPS basis $\left\Vert \bm{\alpha}^{(\bm{q}')}\right\rangle _{\mathcal{Q}}$
with a different phase or radius. For this mapping of $\tilde{\alpha}\rightarrow\alpha^{(q)}$,
on defining $z_{q}=\tilde{\alpha}/\alpha^{(q)}$ , the expansion coefficient
in the single-mode case has a rational function expression:

\begin{equation}
\psi_{q}\left(\tilde{\alpha}\right)=\frac{1}{d}\sum_{n=0}^{n_{m}}z_{q}^{n}=\frac{1}{d}\left[\frac{1-z_{q}^{d}\,}{1-z_{q}}\right]\,.\label{eq:coherent CPS re-expansion}
\end{equation}

Alternatively, using normalized states with $\Psi_{\bm{q}}=g_{\mathcal{Q}}\psi_{\bm{q}}$so
that$\left|\psi\right\rangle =\sum_{\bm{q}}\Psi_{\bm{q}}\left|\bm{\alpha}^{(\bm{q})}\right\rangle _{\mathcal{Q}}\,,$
one would obtain:
\begin{equation}
\Psi_{\bm{q}}\left(\tilde{\alpha}\right)=\frac{g_{\mathcal{Q}}\left(\alpha\right)}{g_{\mathcal{Q}}\left(\tilde{\alpha}\right)d}\left[\frac{1-z_{q}^{d}\,}{1-z_{q}}\right]\,.
\end{equation}

\section{\label{sec:Projected-state-identites}Projected state identities}

To use phase-space distributions in a calculation, one needs operator
identities. These are employed to compute observables and calculate
dynamical equations of motion. Here identities are obtained for operators
acting on projected coherent states. For a projector $\hat{\mathcal{P}}$,
there are invariant operators which satisfy $\hat{O}_{I}=\mathcal{P}\hat{O}_{I}\mathcal{P}$.
Such operators have straightforward identities, and there are identities
available in more general cases as well. 

\subsection{General projected identities}

Many operators acting on coherent states correspond to simple differential
forms. These differential identities are expressed as
\begin{eqnarray}
\hat{O}\left\Vert \bm{\alpha}\right\rangle  & = & \mathcal{D}\left(\bm{\alpha}\right)\left\Vert \bm{\alpha}\right\rangle \,\label{eq:unnormalized-operator-identity}
\end{eqnarray}
where $\mathcal{D}\left(\bm{\alpha}\right)$ is a differential operator
of form $\mathcal{D}\left(\bm{\alpha}\right)=\sum_{n}D^{(n)}\left(\bm{\alpha}\right)\partial{}_{i_{1}}\ldots\partial{}_{i_{n}}$,
with the notation that:
\begin{equation}
\partial_{i}\equiv\frac{\partial}{\partial\alpha_{i}}\,.
\end{equation}

In the projected case, there are differential identities available.
If $\hat{O}_{\mathcal{P}}$ is a projected operator, one may define
in analogy to (\ref{eq:unnormalized-operator-identity}) that: 

\begin{eqnarray}
\hat{O}_{\mathcal{P}}\left\Vert \bm{\alpha}\right\rangle _{\mathcal{P}} & = & \mathcal{D}_{\mathcal{P}}\left(\bm{\alpha}\right)\left\Vert \bm{\alpha}\right\rangle _{\mathcal{P}}\,.\label{eq:unnormalized-operator-identity-projected}
\end{eqnarray}

Detailed examples of these are given below; where they exist, they
are very similar to those for the standard coherent states.

If the operator $\hat{O}$ is invariant, then $\mathcal{P}$ commutes
both with $\hat{O}$ and with any differential operator $\mathcal{D}\left(\bm{\alpha}\right)$,
so that there is always one differential identity available that matches
the non-projected case:

\begin{equation}
\hat{O}\left\Vert \bm{\alpha}\right\rangle _{\mathcal{P}}=\mathcal{D}\left(\bm{\alpha}\right)\left\Vert \bm{\alpha}\right\rangle _{\mathcal{P}}\,.\label{eq:InvariantStateIdentitty}
\end{equation}

For projections onto normalized coherent states, the invariant operator
identities depend on the projection, because of the changed normalization
factor. On re-arranging the ordering of differential terms, this becomes:
\begin{equation}
\hat{O}\left|\bm{\alpha}\right\rangle _{\mathcal{P}}=\left[\mathcal{D}+a_{\mathcal{P}}\right]\left|\bm{\alpha}\right\rangle _{\mathcal{P}}\,.\label{eq:InvariantNormStateIdentitty}
\end{equation}
where the c-number term $a_{\mathcal{P}}^{(\psi)}$ for state identities
is obtained using a differential commutator:
\begin{equation}
a_{\mathcal{P}}=g_{\mathcal{P}}^{-1}\left[\mathcal{D},g_{\mathcal{P}}\right]\,.\label{eq:state-projected-identity}
\end{equation}
The differential commutator is defined in the usual way as $\left[\mathcal{D},g\left(\bm{\alpha}\right)\right]\equiv\left[\mathcal{D}g\left(\bm{\alpha}\right)-g\left(\bm{\alpha}\right)\mathcal{D}\right]$,
with differential operators acting on all functions to the right.

\subsection{Number-conserving operator identities}

To illustrate this, consider total number conserving operators acting
on total number-projected coherent states. Similar identities hold
as for the usual coherent states, since these operators are invariant.

For example, for the number operator $\hat{n}=\hat{a}^{\dagger}\hat{a}$
acting an unprojected coherent state, $\hat{n}\left\Vert \alpha\right\rangle =\alpha\partial_{\alpha}\left\Vert \alpha\right\rangle \,$.
As expected from (\ref{eq:InvariantStateIdentitty}), for an invariant
operator this identity is the same in the number-projected case: 
\begin{eqnarray}
\hat{n}\left\Vert \alpha\right\rangle _{\mathcal{N}} & = & \alpha\partial_{\alpha}\left\Vert \alpha\right\rangle _{\mathcal{\mathcal{N}}}\,.
\end{eqnarray}

However, there  is a change for \emph{normalized} coherent states.
Including normalization, the original non-projected identity is
\begin{equation}
\hat{n}\left|\alpha\right\rangle =\left[\alpha\partial_{\alpha}+\frac{1}{2}\left|\alpha\right|^{2}\right]\left\Vert \alpha\right\rangle \,,\label{eq:number-coherent-state-identity}
\end{equation}
and on applying Eq (\ref{eq:state-projected-identity}) to the multi-mode
case, one obtains the general result:
\begin{eqnarray}
\hat{n}_{ij}\left|\bm{\alpha}\right\rangle _{\mathcal{\mathcal{N}}} & = & \alpha_{j}\left[\partial_{\alpha_{i}}+\partial_{\alpha_{i}}\ln g_{\mathcal{\mathcal{N}}}\right]\left|\bm{\alpha}\right\rangle _{\mathcal{\mathcal{N}}}\,.
\end{eqnarray}
This is generally different to the identity of (\ref{eq:number-coherent-state-identity}),
since the normalization $g_{\mathcal{N}}$ depends on the projection. 

\subsection{Differential qudit identities}

For qudit operator identities, the single mode number identities are
invariant, so that if $\hat{n}=\hat{a}^{\dagger}\hat{a}$,
\begin{equation}
\hat{n}\left\Vert \alpha\right\rangle _{\mathcal{Q}}=\alpha\partial_{\alpha}\left\Vert \alpha\right\rangle _{\mathcal{Q}}\,,
\end{equation}
and for a normalized qudit coherent state projection,
\begin{equation}
\hat{n}\left|\alpha\right\rangle _{Q}=\alpha\left[\partial_{\alpha}+\partial_{\alpha}\ln g_{\mathcal{\mathcal{Q}}}\right]\left|\alpha\right\rangle _{Q}\,.
\end{equation}

In the CPS case, the resulting normalized identities are similar to
the usual coherent state identities, except for an additional correction
term that vanishes for $n_{0}\ll\left|\alpha\right|^{2}\ll n_{m}$,
since: 
\begin{equation}
\partial_{\alpha}\ln g_{\mathcal{\mathcal{Q}}}=\frac{\alpha^{*}}{2}\left[1+\left(g_{n_{0}-1}^{2}-g_{n_{m}}^{2}\right)g_{Q}^{-2}\right]\,.
\end{equation}

For annihilation or creation operators, which are non-invariant, one
can still define projected operator identities following (\ref{eq:unnormalized-operator-identity-projected})
. In a single-mode CPS, the action of the creation operator is:
\begin{equation}
\hat{a}^{\dagger}\left\Vert \alpha\right\rangle _{\mathcal{\mathcal{Q}}}=\sum_{n=n_{0}}^{n_{m}}\frac{\left(n+1\right)\alpha^{n}}{\sqrt{\left(n+1\right)!}}\left|n+1\right\rangle ,
\end{equation}
 which leads to the differential identity:
\begin{equation}
\hat{a}^{\dagger}\left\Vert \alpha\right\rangle _{\mathcal{\mathcal{Q}}}=\partial_{\alpha}\left[\left\Vert \alpha\right\rangle _{\mathcal{\mathcal{\mathcal{\mathcal{Q}}}}}+\left|\alpha\right\rangle _{n_{m}+1}-\left|\alpha\right\rangle _{n_{0}}\right]\,.
\end{equation}

Therefore, on projection, one obtains a simple closed identity in
the special case of $n_{0}=0$: 
\begin{equation}
\hat{a}_{\mathcal{\mathcal{\mathcal{Q}}}}^{\dagger}\left\Vert \alpha\right\rangle _{\mathcal{\mathcal{\mathcal{Q}}}}=\partial_{\alpha}\left\Vert \alpha\right\rangle _{\mathcal{\mathcal{\mathcal{Q}}}}\,.
\end{equation}
For the annihilation operator, the corresponding result is no longer
simply expressed in terms of the original projected basis, since
\begin{equation}
\hat{a}\left\Vert \alpha\right\rangle _{\mathcal{\mathcal{\mathcal{\mathcal{Q}}}}}=\alpha\left[\left\Vert \alpha\right\rangle _{\mathcal{\mathcal{\mathcal{\mathcal{Q}}}}}+\left|\alpha\right\rangle _{n_{0}-1}-\left|\alpha\right\rangle _{n_{m}}\right]\,.
\end{equation}

On projecting the operator, one obtains a closed identity which is
asymptotically valid when the tails of the distribution $g_{n}$ are
negligible, so that
\begin{equation}
\hat{a}_{\mathcal{\mathcal{\mathcal{Q}}}}\left\Vert \alpha\right\rangle _{\mathcal{\mathcal{\mathcal{Q}}}}\cong\alpha\left\Vert \alpha\right\rangle _{\mathcal{\mathcal{\mathcal{Q}}}}\,.
\end{equation}
If $n_{0}=0$, the lower limit correction vanishes. The high-$n$
term becomes negligible exponentially fast as $n_{m}$ is increased
due to the Gaussian cut-off at large $n_{m}.$ Another route is to
use a small coherent radius, $r=\left|\alpha\right|\ll1,$ as the
starting point of a CPS mapping, which can also strongly suppress
the high-$n$ residual term.

\subsection{CPS matrix identities}

For the CPS case, there are general matrix identities with the form
\begin{equation}
\hat{O}_{\mathcal{Q}}\left|\bm{\alpha}^{(\bm{q})}\right\rangle _{\mathcal{Q}}=\sum_{\bm{q}'}\mathcal{O}_{\bm{q}',\bm{q}}\left(\bm{\alpha}\right)\left|\bm{\alpha}^{(\bm{q}')}\right\rangle _{\mathcal{Q}}.\label{eq:CPS matrix-identities}
\end{equation}

Utilizing the identity expansion (\ref{eq:identity-scaled-number}),
and introducing the matrix $O_{n'n}(\alpha)=g_{n'}^{-2}\left\langle \text{\ensuremath{\alpha}}\right\Vert _{n'}\hat{O}\left\Vert \alpha\right\rangle _{n}$,
in the single-mode case this becomes: 
\begin{equation}
\hat{O}_{\mathcal{Q}}\left|\alpha^{(q)}\right\rangle _{\mathcal{Q}}=g_{\mathcal{Q}}^{-1}\sum_{n,n'=0}^{d-1}e^{iqn\phi}O_{n'n}(\alpha)\left\Vert \alpha\right\rangle _{n'}\,.
\end{equation}
Therefore, on inverting the mapping using the discrete Fourier transform
theorem,
\begin{eqnarray}
\hat{O}_{\mathcal{Q}}\left|\alpha^{(q)}\right\rangle _{\mathcal{Q}} & = & \sum_{q'=0}^{d-1}\mathcal{O}_{q'q}\left(\alpha\right)\left|\alpha^{(q')}\right\rangle _{\mathcal{Q}}\,.
\end{eqnarray}
where the matrix $\mathcal{O}_{q'q}\left(\alpha\right)$ is: 
\begin{equation}
\mathcal{O}_{q'q}\left(\alpha\right)=\frac{1}{d}\sum_{n,n'=0}^{d-1}O_{n'n}\left(\alpha\right)e^{i(qn-q'n')\phi}\,.
\end{equation}
One can therefore evaluate any pure state expectation value as: 
\begin{equation}
\left\langle \hat{O}\right\rangle =\sum_{\bm{q}_{i}}\Psi_{\bm{q}_{1}}^{*}M_{\bm{q}_{1}\bm{q}_{2}}\mathcal{O}_{\bm{q}_{2}\bm{q}_{3}}\Psi_{\bm{q}_{3}}=\bm{\Psi}^{\dagger}\bm{M}\bm{\mathcal{O}}\bm{\Psi}\,.
\end{equation}
where the inner-product matrix $M_{\bm{q}_{1}\bm{q}_{2}}=\left\langle \bm{\alpha}^{(\bm{q}_{1})}\right.\left|\bm{\alpha}^{(\bm{q}_{2})}\right\rangle $
takes account of the non-orthogonality of the CPS basis. The choice
of matrix or differential identity that is used depends on the requirements
of a particular calculation. Differential identities are most useful
for linear evolution, while matrix identities have more general applicability.

\subsection{Hamiltonian evolution with CPS identities\label{subsec:Qudit-discrete-identities}}

As well as the differential results given above, the discrete identities
for coherent phase states can be used to calculate the time evolution
of a projection invariant state $\left|\psi(t)\right\rangle $. Firstly,
the state is expanded in terms of the non-orthogonal CPS states, in
the form
\begin{equation}
\left|\psi(t)\right\rangle =\sum_{q=0}^{d-1}\Psi_{q}(t)\left|\alpha^{(q)}\right\rangle _{\mathcal{Q}}\,.
\end{equation}

Using a CPS mapping of $\hat{H}/\hbar\rightarrow\bm{\mathcal{H}}$,
the Hamiltonian evolution of the expansions is given by:
\begin{align}
i\frac{\partial}{\partial t}\left|\psi(t)\right\rangle  & =\sum_{q=0}^{d-1}\mathcal{H}_{q'q}\Psi_{q}(t)\left|\alpha^{(q)}\right\rangle _{\mathcal{Q}}.
\end{align}

To obtain the identities, for $\mathcal{\bm{H}}$, the zero-based
CPS case will be used here, with $n_{0}=0$. 

The CPS matrices can be evaluated in common cases, by setting $z=\exp\left(i\left(q'-q\right)\phi\right)$,
as follows:
\begin{eqnarray}
\hat{a}\rightarrow\mathcal{O}_{q'q}^{[a]}\left(\alpha\right) & = & \alpha^{(q)}\left[\delta_{qq'}-1/d\right]\label{eq:general-identities}\\
\hat{n}^{k}\rightarrow\mathcal{O}_{q'q}^{[k]} & = & \frac{1}{d}\sum_{n=0}^{d-1}n^{k}z^{n}\nonumber \\
\hat{a}^{\dagger}\rightarrow\mathcal{O}_{q'q}^{[a^{\dagger}]}\left(\alpha\right) & = & \frac{1}{\alpha^{(q)}}\mathcal{O}_{q'q}^{[1]}.\nonumber 
\end{eqnarray}
The projected identities are all matrix operations acting on the coherent
phase states, which do not depend on the amplitude $\alpha$ for invariant
operators. The identities for $\hat{n}^{k}$ involve sums over $z$
that are well-known :

\begin{align}
\mathcal{O}_{q'q}^{[1]} & =\frac{z-z^{d}\left[d-z\left(d-1\right)\right]}{d\left(1-z\right)^{2}}\nonumber \\
 & =(d-1)/2\,\,\,\,\,\,\,\,\,\,\,\,\,(q=q')\nonumber \\
\mathcal{O}_{q'q}^{[2]} & =\frac{z+z^{2}-z^{d}\left[d-z\left(d-1\right)\right]^{2}-z^{d+1}}{d\left(1-z\right)^{3}}\nonumber \\
 & =(d-1)(2d-1)/6\,\,\,\,\,\,\,\,\,\,\,(q=q')\,.\label{eq:nonlinearpropagationmatrix}
\end{align}

The nonlinear matrices grow with the cut-off $d$, although the matrix
corresponding to $\hat{n}^{2}-n_{m}\hat{n}$ has no terms except diagonals
that grow with $d:$

\begin{align}
O_{q'q}^{(2)}-n_{m}O_{q'q}^{(1)} & =z\frac{z^{d}(d-2)-d(z^{d-1}-z+1)+2}{\left(1-z\right)^{3}}\,\nonumber \\
 & =(d-1)(2-d)/6\,\,\,\,\,\,\,\,\,\,\,\,\,(q=q')\,.
\end{align}

After applying these identities, the time evolution of the vector
of expansion coefficients $\bm{\Psi}$ is given by a unitary matrix
$\bm{\mathcal{U}}(t)=\exp\left(-i\bm{\mathcal{H}}t\right)$:
\begin{equation}
\bm{\Psi}(t)=\bm{\mathcal{U}}(t)\bm{\Psi}(0).
\end{equation}
This is a finite matrix equation, and directly uses the discrete Fourier
transform identity (\ref{eq:inverse-DFT-phase-coherent}). 

Due to the linearity of quantum mechanics, the unitary evolution matrix
$\bm{\mathcal{U}}(t)$ is independent of the initial state. This means
that once $\bm{\mathcal{U}}(t)$ is known, the time evolution of any
state within the projected manifold is immediately obtainable. 

\subsection{Hybrid pure-state evolution}

From Eq (\ref{eq:nonlinearpropagationmatrix}), there is an unexpected
invariance for Hamiltonians of form $\hat{n}^{k}$, which only depend
on the particle number. In these cases, $\bm{\mathcal{U}}(t)$ is
independent of the amplitude $\alpha$. This can be used to obtain
a novel type of factorization of the dynamical evolution. Consider
a Hamiltonian with both a possibly time-dependent linear term $\hat{H}^{(0)}(t)=\hbar\omega_{ij}(t)\hat{a}_{i}^{\dagger}\hat{a}_{j}$
and a nonlinear term $\hat{H}^{(n)}$ which only has polynomial terms
in $\hat{n}^{k}$, so that: 
\begin{equation}
\hat{H}=\hat{H}^{(0)}(t)+\hat{H}^{(n)}\,.
\end{equation}

One way to treat such problems is to use the interaction picture of
quantum mechanics, in which operators evolve according to one part
of the Hamiltonian and wavefunctions according to another. A variant
of this approach is possible with projected coherent states. With
this hybrid evolution approach, the base coherent state amplitude
$\alpha$ evolves according to linear differential identities, while
the phase indices $q$ evolve according to nonlinear discrete identities
which are independent of $\alpha$. 

In greater detail, one expands using both the discrete indices and
the base amplitude, so that using Einstein summation over repeated
$\bm{q}$ indices:

\begin{equation}
\left|\psi(t)\right\rangle =\int\Psi_{\bm{q}}(\bm{\alpha},t)\left|\bm{\alpha}^{(\bm{q})}\right\rangle _{\mathcal{Q}}d\bm{\alpha}\,.
\end{equation}

Under time-evolution,
\begin{equation}
i\hbar\frac{\partial}{\partial t}\left|\psi(t)\right\rangle =\int\Psi_{\bm{q}}(\bm{\alpha},t)\hat{H}\left|\bm{\alpha}^{(\bm{q})}\right\rangle _{\mathcal{Q}}d\bm{\alpha}\,.
\end{equation}

Next, one uses identities so that $\hat{H}^{(0)}(t)/\hbar\rightarrow\mathcal{D}\left(\bm{\alpha},t\right)$,
$\hat{H}^{(n)}/\hbar\rightarrow\bm{\mathcal{H}}^{(n)}$. Provided
partial integration is possible, leading to a partially integrated
operator $\mathcal{D}'\left(\bm{\alpha},t\right)$, the resulting
time-evolution equation is:

\begin{equation}
\dot{\Psi}_{\bm{q}}(\bm{\alpha},t)=-i\left[\mathcal{D}'\left(\bm{\alpha},t\right)+\mathcal{H}_{\bm{q}\bm{q}'}^{(n)}\right]\Psi_{\bm{q}}(\bm{\alpha},t)\,.
\end{equation}

If $\mathcal{H}_{\bm{q}\bm{q}'}^{(n)}$ is independent of $\bm{\alpha}$,
which is true with number-conserving nonlinear Hamiltonians, one may
introduce an integrating factor $\bm{\mathcal{U}}^{(n)}(t)=\exp\left(-i\bm{\mathcal{H}}^{(n)}t\right)$
such that $\Psi_{\bm{q}}(t)=\bm{\mathcal{U}}^{(n)}(t)\Psi_{\bm{q}}^{(n)}(t)$,
$\tilde{\mathcal{D}}'\left(\bm{\alpha},t\right)=\bm{\mathcal{U}}^{(n)\dagger}(t)\mathcal{D}'\left(\bm{\alpha},t\right)\bm{\mathcal{U}}^{(n)}(t)$.
The resulting differential equation is then:
\begin{equation}
\dot{\Psi}_{\bm{q}}^{(n)}(\bm{\alpha},t)=\tilde{\mathcal{D}}'\left(\bm{\alpha},t\right)\Psi_{\bm{q}}^{(n)}(\bm{\alpha},t)\,.\label{eq:hybrid}
\end{equation}

This has the unusual feature that it is the nonlinear term rather
than the linear term that is removed in the interaction picture. For
single mode Hamiltonians the situation is even simpler, since $\tilde{\mathcal{D}}'\left(\bm{\alpha},t\right)=\mathcal{D}'\left(\bm{\alpha},t\right)=i\omega(t)\partial_{\alpha}\alpha$.
In this case the time evolution factorizes completely, such that $\alpha$
can be treated as a characteristic trajectory obeying $\dot{\alpha}(t)=-i\omega\alpha(t)$
. This causes a rotation in the reference amplitude $\alpha$, while
the nonlinear terms independently modify the discrete indices $q$
. 

\section{Projected P-representations}

The generalized P-representation maps a quantum Hilbert space to a
phase-space of double the classical dimensionality \cite{Drummond_Gardiner_PositivePRep,Chaturvedi:1994}.
It provides a natural way to represent mixed states rather than just
pure states, and is especially useful in open and exponentially complex
systems. Yet if just part of a Hilbert space is physically important,
due to a symmetry or other reasons, it is more efficient to only map
the projected part of the space.

For such projected measurements, it is irrelevant whether one projects
the density matrix, the measured operator, or both. Given a density
matrix $\hat{\rho}$ or an operator $\hat{O}$, their projected versions
are $\hat{\rho}_{\mathcal{P}}$ and $\hat{O}_{\mathcal{P}}$ respectively.
Projections of the density matrix allow the treatment of all possible
projected measurements in a unified way. For projected phase-space
representations, suppose that the density matrix $\hat{\rho}$ is
invariant under the projection, i.e.,
\begin{equation}
\hat{\rho}=\hat{\rho}_{\mathcal{P}}=\mathcal{P}\hat{\rho}\mathcal{P}\,.
\end{equation}

This section will develop efficient mappings onto a phase-space for
density matrices of this type, that occupy only a fractional part
of the Hilbert space. The projection types are as in the earlier sections.
Two different types of identity, for normalized or un-normalized representations,
are treated.

\subsection{Generalized P-representations}

It is useful to first review the P-representation phase-space method.
This was originally proposed in a diagonal form \cite{Glauber_1963_P-Rep,Sudarshan_1963_P-Rep},
where it was crucial to the development of coherence and laser theory.
Since this gave singular results for nonclassical states, the generalized
P-representation was introduced for all quantum states \cite{drummond2016quantum},
with the form:

\begin{equation}
\hat{\rho}=\int P\left(\overrightarrow{\bm{\alpha}}\right)\hat{\Lambda}\left(\overrightarrow{\bm{\alpha}}\right)d\mu\,.\label{eq:Projected-P-definition}
\end{equation}
Here $\overrightarrow{\bm{\alpha}}\equiv\left(\bm{\alpha},\bm{\beta}\right)$
denotes a double-dimensional coherent amplitude, and the operator
basis used in the expansion is:

\begin{equation}
\hat{\Lambda}\left(\overrightarrow{\bm{\alpha}}\right)\equiv e^{-\bm{\alpha}\cdot\bm{\beta}}\left\Vert \bm{\alpha}\right\rangle \left\langle \bm{\beta}^{*}\right\Vert =G^{-1}\left(\overrightarrow{\bm{\alpha}}\right)\hat{\lambda}\left(\overrightarrow{\bm{\alpha}}\right).
\end{equation}

The normalizing factor is $G\left(\overrightarrow{\bm{\alpha}}\right)=\exp\left[\bm{\alpha}\cdot\bm{\beta}\right]$,
which reduces to $g^{2}\left(\bm{\alpha}\right)$ as in (\ref{eq:normalized-coherent-state}),
in the diagonal limit of $\bm{\alpha}=\bm{\beta}^{*}$. In positive
P-representations, the integration measure $d\mu$ is a volume measure,
$d\mu=d^{2M}\bm{\alpha}d^{2M}\bm{\beta}$, which has double the classical
dimension. The measure can also be a contour integral, so that $d\mu=d\bm{\alpha}d\bm{\beta}$,
and the distribution function $P\left(\overrightarrow{\bm{\alpha}}\right)$
is complex-valued. There are several existence theorems for the full
Hilbert space \cite{Drummond1980,drummond2014quantum}. 
\begin{itemize}
\item A positive, normally ordered standard positive P-distribution exists
for all quantum states. While this construction is always available,
it is often not the most compact form.
\item A complex P-representation, requiring a contour integral around the
origin, exists for all states with a bounded support in number or
coherent states. This approach can give a very compact distribution.
It is a special case of more general gauge, or complex weighted methods,
that are useful in simulating quantum dynamics \cite{Deuar:2002}.
\end{itemize}
In both cases, the representation gives observable normally-ordered
correlation functions that are classical-like moments of the distribution. 

\subsection{Projected P-representations}

The projected P-representation has a similar construction to the generalized
P-representation, except that it makes use of a projected operator
basis, so that:

\begin{align}
\hat{\rho} & =\int p\left(\overrightarrow{\bm{\alpha}}\right)\hat{\lambda}_{\mathcal{P}}\left(\overrightarrow{\bm{\alpha}}\right)d\mu\,.\label{eq:Projected-P-definition-3}\\
 & =\int P\left(\overrightarrow{\bm{\alpha}}\right)\hat{\Lambda}_{\mathcal{P}}\left(\overrightarrow{\bm{\alpha}}\right)d\mu\,.
\end{align}
Here, the unnormalized projected basis $\hat{\lambda}_{\mathcal{P}}$
is: 
\begin{equation}
\hat{\lambda}_{\mathcal{P}}\left(\overrightarrow{\bm{\alpha}}\right)\equiv\left\Vert \bm{\alpha}\right\rangle _{P}\left\langle \bm{\beta}^{*}\right\Vert _{P}\,.
\end{equation}
and the projected basis $\hat{\Lambda}_{P}$ normalized so that $Tr\left(\hat{\Lambda}_{P}\right)=1$
is 
\begin{equation}
\hat{\Lambda}_{\mathcal{P}}\left(\overrightarrow{\bm{\alpha}}\right)\equiv G_{\mathcal{P}}^{-1}\left(\overrightarrow{\bm{\alpha}}\right)\hat{\lambda}_{\mathcal{P}}\left(\overrightarrow{\bm{\alpha}}\right),
\end{equation}
so that $p\left(\overrightarrow{\bm{\alpha}}\right)=G_{\mathcal{P}}^{-1}\left(\overrightarrow{\bm{\alpha}}\right)P\left(\overrightarrow{\bm{\alpha}}\right).$
The normalization factor, $G_{P}\left(\overrightarrow{\bm{\alpha}}\right)\equiv\left\langle \bm{\beta}^{*}\right\Vert \left.\bm{\alpha}\right\rangle _{\mathcal{P}}$,
is the inner product of two unnormalized projected coherent states,
so that for number projections:
\begin{equation}
G_{\mathcal{N}}\left(\overrightarrow{\bm{\alpha}}\right)=\sum_{\bm{n}\in S}\prod_{i}\frac{\left(\beta_{i}\alpha_{i}\right)^{n_{i}}}{n_{i}!}\,.
\end{equation}

The projected P-distribution is generally different to the standard
P-distribution, even for an identical density matrix. The reason for
this is that the operator basis has a changed normalization. Unless
one takes the limit of $G_{\mathcal{P}}\left(\overrightarrow{\bm{\alpha}}\right)\rightarrow1$,
the normalized projected basis is not just a projection of the normalized
basis, i.e., in general 
\begin{equation}
\hat{\Lambda}_{\mathcal{P}}\left(\overrightarrow{\bm{\alpha}}\right)\neq\mathcal{P}\hat{\Lambda}\left(\overrightarrow{\bm{\alpha}}\right)\mathcal{P}\,.
\end{equation}
This is because the full operator basis extends over the entire Hilbert
space, so it has a different normalization. For a projection invariant
density matrix, the unprojected distribution has to ensure that the
expansion in coherent states has no contribution from the additional
part of the Hilbert space. This is not required of the projected distribution.
One can obtain a projected distribution from an unprojected one, by
first projecting then renormalizing it.

The expansions are identical in the limit $\left|\bm{\alpha}\right|\rightarrow0$,
$\left|\bm{\beta}\right|\rightarrow0$, where $G_{\mathcal{P}}\left(\overrightarrow{\bm{\alpha}}\right)\rightarrow1$.
The un-normalized projected operator basis is similar to the unprojected
case, since 
\begin{equation}
\hat{\lambda}_{\mathcal{P}}\left(\overrightarrow{\bm{\alpha}}\right)=\mathcal{P}\hat{\lambda}\left(\overrightarrow{\bm{\alpha}}\right)\mathcal{P}\,.\label{eq:projection-identity}
\end{equation}
However, this is not the case for the normalized basis, due to the
additional normalizing factor. 

\subsection{Existence of number-projected P-distributions}

A similar method can be to obtain existence theorems for the number-projected
case as for the complex P-representation \cite{Drummond_Gardiner_PositivePRep}.
Recalling the definition of $\left\Vert \bm{\alpha}\right\rangle _{\bm{n}}$,
and using Cauchy's theorem, a contour integral simply projects out
the appropriate number state. On substitution into the general definition
of the projected phase-space representation, an un-normalized P-representation
is obtained where: 
\begin{eqnarray}
p\left(\overrightarrow{\bm{\alpha}}\right) & = & \sum_{\bm{n}\bm{m}\in S}\frac{\rho_{\mathbf{m}\mathbf{n}}}{\left(2\pi i\right)^{2M}}\int d\mu\prod_{j}\frac{\sqrt{n_{j}!m_{j}!}}{\beta_{j}^{n_{j}+1}\alpha_{j}^{m_{j}+1}}\,.\nonumber \\
 &  & \,
\end{eqnarray}

This contour integral method can be used to obtain existence theorems
for any finite state case, regardless of which type of number projection
is used. This gives a continuous rather than a discrete distribution,
which is complex-valued. It is applicable to both normalized and un-normalized
cases, since one can always obtain a normalized distribution through
the relationship that $P=pG_{\mathcal{P}}$. 

There appears to be no corresponding exact construction for projected
positive P-distributions, except in special cases. This is not a significant
issue: one can still calculate observables with complex weights.

\subsection{CPS P-representations }

In the case of coherent phase states, from (\ref{subsec:Coherent-phase-states}),
one can define the measure as a discrete summation over a set $\tilde{S}$
of $d^{2M}$ coherent amplitudes that is complementary to the projected
set $S$. This is similar to a contour integral, except that the distribution
is a sum over delta-functions in a circle, not a continuous integral.
Here, one defines the CPS amplitudes similarly to a wave-function
expansion, so that:
\begin{eqnarray}
\alpha^{(q)} & \equiv & \alpha\exp\left(iq\phi\right),\,\,\,\,\nonumber \\
\beta^{(q)} & \equiv & \beta\exp\left(-iq\phi\right),\,\,\,\,q=0,\ldots d-1.
\end{eqnarray}
This is identical to defining $P_{\mathcal{Q}}\left(\overrightarrow{\bm{\alpha}}\right)$
as a delta-function at each discrete point in phase-space. In the
CPS case, one can introduce a set of basis operators$\hat{\lambda}_{\overrightarrow{\bm{q}}}\left(\overrightarrow{\bm{\alpha}}\right)\equiv\hat{\lambda}_{\mathcal{Q}}\left(\bm{\alpha}^{(\bm{q})},\bm{\beta}^{(\bm{q}')}\right)$,
where $\overrightarrow{\bm{q}}\equiv\left(\bm{q},\bm{q}'\right)$.
The normalization factor is $G_{\overrightarrow{\bm{q}}}\left(\overrightarrow{\bm{\alpha}}\right)=G_{\mathcal{Q}}\left(\bm{\alpha}^{(\bm{q})},\bm{\beta}^{(\bm{q}')}\right)$
and the corresponding normalized basis is defined as:
\begin{align*}
\hat{\Lambda}_{\overrightarrow{\bm{q}}}\left(\overrightarrow{\bm{\alpha}}\right) & \equiv\hat{\Lambda}_{\mathcal{Q}}\left(\bm{\alpha}^{(\bm{q})},\bm{\beta}^{(\bm{q}')}\right)\,.\\
 & =\hat{\lambda}_{\overrightarrow{\bm{q}}}\left(\overrightarrow{\bm{\alpha}}\right)/G_{\overrightarrow{\bm{q}}}\left(\overrightarrow{\bm{\alpha}}\right)\,.
\end{align*}
The representation is then written as:
\begin{align}
\hat{\rho} & =\sum_{\overrightarrow{\bm{q}}}p_{\overrightarrow{\bm{q}}}\left(\overrightarrow{\bm{\alpha}}\right)\hat{\lambda}_{\overrightarrow{\bm{q}}}\left(\overrightarrow{\bm{\alpha}}\right)\,\nonumber \\
 & =\sum_{\overrightarrow{\bm{q}}}P_{\overrightarrow{\bm{q}}}\left(\overrightarrow{\bm{\alpha}}\right)\hat{\Lambda}_{\overrightarrow{\bm{q}}}\left(\overrightarrow{\bm{\alpha}}\right)\,,
\end{align}

The single-mode normalization coefficient is given explicitly by:
\begin{equation}
G_{\mathcal{Q}}\left(\alpha,\beta\right)=\sum_{n=n_{0}}^{n_{m}}\frac{\left(\beta\alpha\right)^{n}}{n!}\,.
\end{equation}

CPS P-representations for projected density matrices using coherent
phase states can be found following the methods given in Section (\ref{subsec:Qudit-bosonic-coherent}).
As described in Section (\ref{subsec:Coherent-phase-states}), this
method can be utilized if the set of occupation numbers for mode $j$
has the form $n_{j}=n_{0},\ldots n_{m}$, where $n_{m}=d+n_{0}-1$.
An example is for $n_{0}=0$, so that the first state is the vacuum
state. With this expansion, a unique CPS P-function $p_{\mathbf{q},\mathbf{q}'}$
always exists given $\rho_{\mathbf{m}\mathbf{n}}=\left\langle \bm{m}\right|\widehat{\rho}_{\mathcal{Q}}\left|\bm{n}\right\rangle $,
where:
\begin{equation}
p_{\mathbf{q},\mathbf{q}'}=\sum_{\bm{n},\bm{m}\in S}\rho_{\mathbf{m}\mathbf{n}}\prod_{j}\frac{\sqrt{n_{j}!m_{j}!}}{d^{2}\left(\alpha_{j}^{(q_{j})}\right)^{m_{j}}\left(\beta_{j}^{(q'_{j})}\right)^{n_{j}}}\,.\label{eq:existence-1-1-1}
\end{equation}
Consequently, a normalized $P_{\mathbf{q},\mathbf{q}'}=p_{\mathbf{q},\mathbf{q}'}G_{\mathbf{q},\mathbf{q}'}$.
This can be verified as a solution by inserting this distribution
into the expansion of the density matrix, noting that, from the properties
of the discrete Fourier transform, 
\begin{equation}
\frac{1}{d^{M}}\sum_{\mathbf{k}}e^{i\bm{q}\cdot\left(\bm{n}-\bm{m}\right)}=\delta_{\bm{n}-\bm{m}}\,.
\end{equation}

This approach is independent of the base amplitudes $\alpha$and $\beta$.
A second method of expansion is that if one representation is known
that uses a coherent projector on a different basis with $\hat{\lambda}_{Q}\left(\tilde{\bm{\alpha}},\tilde{\bm{\beta}}\right)$,
one can map $\tilde{\alpha},\tilde{\beta}\rightarrow\alpha^{(q)},\beta^{(q)}$
using Eq (\ref{eq:coherent CPS re-expansion}), so that:
\begin{equation}
\hat{\lambda}_{Q}\left(\tilde{\bm{\alpha}},\tilde{\bm{\beta}}\right)=\sum_{\overrightarrow{q}}\tilde{p}_{\overrightarrow{q}}\hat{\lambda}_{\overrightarrow{q}}\,.
\end{equation}
 This allows the transformation of any projected P-function into another
one that uses another basis set, even with a different radius. With
the usual, unprojected complex P-representation, this type of transformation
is known when the contour radius is increased, allowing the use of
Cauchy's theorem \cite{Drummond1980}. In the projected case, any
basis element $\hat{\lambda}_{Q}\left(\tilde{\bm{\alpha}},\tilde{\bm{\beta}}\right)$
can be re-expanded using a CPS basis $\hat{\lambda}_{\overrightarrow{\bm{q}}}$.
The expansion coefficient is similar to the corresponding CPS expression
in Eq (\ref{eq:coherent CPS re-expansion}). Defining $z_{q}=\tilde{\alpha}/\alpha^{(q)}$,
$\zeta_{q}=\tilde{\beta}/\beta^{(q)}$, the coefficient is:
\begin{equation}
\tilde{p}_{\overrightarrow{q}}=\frac{1}{d^{2}}\left[\frac{1-z_{q}^{d}\,}{1-z_{q}}\right]\left[\frac{1-\zeta_{q'}^{d}\,}{1-\zeta_{q'}}\right]\,.\label{eq:coherent Pfunction CPS reprojection}
\end{equation}
 A consequence of this construction is that if necessary, one can
take the small radius limit of $\alpha_{j},\beta_{j}\equiv r\rightarrow0$.
In this limit, the normalizing factor is unity, i.e., $G\rightarrow1$.
This means that the projected distribution is identical to an unprojected
complex P-distribution obtained as a sum of delta-functions for the
same contour, even though the basis is changed. This feature is useful
in calculations \cite{Opanchuk2016-quantum} for photonic networks. 

\subsection{Unnormalized operator identities}

Applications of phase-space representations to quantum physics depend
on the existence of many operator identities, which are closely related
to the coherent state differential identities of Section (\ref{sec:Projected-state-identites}).
Operators acting on $\hat{\Lambda}$ that correspond to differential
operators acting on $P$ allow evaluation of mean expectation values
for the quantum density matrix, as well dynamical evolution equations
for the distribution. This first requires knowledge of a the operator
identities. 

In the non-projected case, one obtains:
\begin{align}
\hat{a}\hat{\lambda} & =\alpha\hat{\lambda}\,.\nonumber \\
\hat{a}^{\dagger}\hat{\lambda} & =\partial_{\alpha}\hat{\lambda}\nonumber \\
\hat{n}\hat{\lambda} & =\alpha\partial_{\alpha}\hat{\lambda}\,.\label{unprojected-operator-identities}
\end{align}
 The consequence of this is that any normally ordered correlation
function is easily evaluated, giving

\begin{equation}
\left\langle a_{m_{1}}^{\dagger}\ldots a_{m_{n}}\right\rangle =\int\beta_{m_{1}}\ldots\alpha_{m_{n}}P\left(\overrightarrow{\bm{\alpha}}\right)d\mu\,.\label{eq:nonclassical correlation}
\end{equation}

What are the corresponding operator identities in the projected case?
For unnormalized P-representations, the identities have an identical
structure to that of \ref{eq:unnormalized-operator-identity-projected},
so that for certain operators $\hat{O}$, 
\begin{eqnarray}
\hat{O}_{\mathcal{P}}\hat{\lambda}_{\mathcal{P}} & = & \mathcal{D}_{\mathcal{P}}\left(\bm{\alpha}\right)\hat{\lambda}_{\mathcal{P}}\nonumber \\
\hat{\lambda}_{\mathcal{P}}\hat{O}_{\mathcal{P}}^{\dagger} & = & \mathcal{D}_{\mathcal{P}}\left(\bm{\beta}\right)\hat{\lambda}_{\mathcal{P}}\,.\label{eq:basis-identities}
\end{eqnarray}

For invariant operators that commute with $\mathcal{P}$, the projection
of an operator product $\hat{O}\hat{\lambda}$ or $\hat{\lambda}\hat{O}$
has a differential identity on $\hat{\lambda}_{\mathcal{P}}$, since:
\begin{equation}
\mathcal{P}\hat{O}\hat{\lambda}\mathcal{P}=\mathcal{P}\mathcal{D}\hat{\lambda}\mathcal{P}=\mathcal{D}\hat{\lambda}_{\mathcal{P}}.
\end{equation}
 Hence, given an operator $\hat{O}$ that commutes with the projector
$\mathcal{P}$, the basis identities of \ref{unprojected-operator-identities}
must follow in the projected case as well. Typical examples are identical
to those given for projected coherent states, so for example, 
\begin{align}
\hat{n}\hat{\lambda}_{\mathcal{N}} & =\alpha\partial_{\alpha}\hat{\lambda}_{\mathcal{N}}\,.
\end{align}

For the CPS case, the general matrix identities of (\ref{eq:CPS matrix-identities})
are directly applicable, giving the result that:
\begin{align}
\hat{O}_{\mathcal{Q}}\hat{\lambda}_{\overrightarrow{\bm{q}}} & =\sum_{\bm{q}"}\mathcal{O}_{\bm{q}",\bm{q}}\left(\bm{\alpha}\right)\hat{\lambda}_{\bm{q}",\bm{q}'}\nonumber \\
\hat{\lambda}_{\overrightarrow{\bm{q}}}\hat{O}_{\mathcal{Q}}^{\dagger} & =\sum_{\bm{q}'}\mathcal{O}_{\bm{q}",\bm{q}',}^{*}\left(\bm{\beta}^{*}\right)\hat{\lambda}_{\bm{q},\bm{q}"}\,.
\end{align}

\subsection{Normalized operator identities}

The corresponding identities for the normalized basis operators are
modified, since the normalizing terms are different after projection.
Thus, one obtains:
\begin{equation}
\hat{O}\hat{\Lambda}_{\mathcal{P}}=G_{\mathcal{P}}\hat{O}\hat{\lambda}_{\mathcal{P}}=G_{\mathcal{P}}\mathcal{D}\hat{\lambda}_{\mathcal{P}}
\end{equation}

On re-arranging the ordering of differential terms, this becomes:
\begin{eqnarray}
\hat{O}\hat{\Lambda}_{\mathcal{P}} & = & \left[\mathcal{D}\left(\bm{\alpha}\right)+\mathcal{A}\left(\overrightarrow{\bm{\alpha}}\right)\right]\hat{\Lambda}_{\mathcal{P}}\nonumber \\
\hat{\Lambda}_{\mathcal{P}}\hat{O}^{\dagger} & = & \left[\mathcal{D}\left(\bm{\beta}\right)+\mathcal{A}\left(\bm{\beta},\bm{\alpha}\right)\right]\hat{\Lambda}_{\mathcal{P}}
\end{eqnarray}
where the c-number terms $\mathcal{A}$ for are given in terms of
a differential commutator. These are similar to the coherent state
identities, except that the normalizing factor is different:
\begin{equation}
\mathcal{A}=G_{\mathcal{P}}^{-1}\left[\mathcal{D},G_{\mathcal{P}}\right]\,.
\end{equation}

It is also possible to obtain identities for more general operator
products that occur in master equations. These are of the form $\hat{O}_{1}\hat{\rho}\hat{O}_{2}$,
which are invariant provided that:
\begin{equation}
\mathcal{P}\hat{O}_{1}\hat{\rho}\hat{O}_{2}\mathcal{P}=\hat{O}_{1}\mathcal{P}\hat{\rho}\mathcal{P}\hat{O}_{2}\,
\end{equation}
However, the corresponding identities are best worked out on a case-by
case basis.

Unlike the quantum state matrix identities, which do not depend on
the normalization, the matrix operator identities for the CPS basis
change when normalized P-representations are considered. The reason
for this is that the normalization factors are state-dependent. As
a a result, the above matrix identities are modified, and become:
\begin{align}
\hat{O}_{\mathcal{Q}}\hat{\Lambda}_{\overrightarrow{\bm{q}}} & =G_{\overrightarrow{\bm{q}}}^{-1}\sum_{\bm{q}"}G_{\bm{q}",\bm{q}'}\mathcal{O}_{\bm{q}",\bm{q}}\left(\bm{\alpha}\right)\hat{\Lambda}_{\bm{q}",\bm{q}'}\nonumber \\
\hat{\Lambda}_{\overrightarrow{\bm{q}}}\hat{O}_{\mathcal{Q}}^{\dagger} & =G_{\overrightarrow{\bm{q}}}^{-1}\sum_{\bm{q}'}G_{\bm{q},\bm{q}"}\mathcal{O}_{\bm{q}",\bm{q}',}^{*}\left(\bm{\beta}^{*}\right)\hat{\lambda}_{\bm{q},\bm{q}"}\,.
\end{align}

\paragraph{Example: the number operator}

As an example, for the single-mode number operator $\hat{n}_{i}=\hat{a}_{i}^{\dagger}\hat{a}_{i}$,
one obtains the differential operator
\begin{equation}
\hat{n}_{i}\left\Vert \bm{\alpha}\right\rangle _{\mathcal{\mathcal{N}}}\left\langle \bm{\beta}^{*}\right\Vert _{\mathcal{\mathcal{N}}}=\alpha_{i}\partial_{i}\left\Vert \bm{\alpha}\right\rangle _{\mathcal{\mathcal{N}}}\left\langle \bm{\beta}^{*}\right\Vert _{\mathcal{\mathcal{N}}}\,.
\end{equation}

Consequently, in the single mode case, with the normalization factor
included:
\begin{equation}
\hat{n}\hat{\Lambda}_{\mathcal{\mathcal{\mathcal{N}}}}=\alpha\left[\partial_{\alpha}+\left(\partial_{\alpha}\ln G_{\mathcal{\mathcal{N}}}\right)\right]\hat{\Lambda}_{\mathcal{\mathcal{N}}}=\left[\mathcal{D}+\mathcal{A}\right]\hat{\Lambda}_{\mathcal{\mathcal{N}}}\,.
\end{equation}
 Here the notation $\left(\partial_{\alpha}\ln G_{\mathcal{\mathcal{N}}}\right)\equiv\mathcal{A}$
indicates that the derivative only acts on the term in the brackets.
For an ensemble average, it follows that: 
\begin{eqnarray}
\left\langle \hat{n}\right\rangle  & = & Tr\left(\int P_{\mathcal{\mathcal{N}}}\left(\alpha,\beta\right)\left[\mathcal{D}+\mathcal{A}\right]\hat{\Lambda}_{\mathcal{\mathcal{N}}}d\mu\right)\,.
\end{eqnarray}

On partial integration and introducing $\mathcal{D}'=\partial_{\alpha}\alpha$
as the re-ordered version of $\mathcal{D}$, after partial integration,
together with the condition that $Tr\left(\hat{\Lambda}_{\mathcal{\mathcal{N}}}\right)=1$,
one obtains:

\begin{eqnarray}
\left\langle \hat{n}\right\rangle  & = & \int\left[\mathcal{A}-\mathcal{D}'\right]P_{\mathcal{\mathcal{N}}}\left(\alpha,\beta\right)d\mu\,.
\end{eqnarray}
Since the integral of a total derivative vanishes due to the boundary
conditions, the final result is a simple moment over the distribution:

\begin{eqnarray}
\left\langle \hat{n}\right\rangle  & = & \int\mathcal{A}P_{\mathcal{\mathcal{N}}}\left(\alpha,\beta\right)d\mu\,.
\end{eqnarray}

In the CPS case, $\mathcal{A}=\alpha\partial_{\alpha}\ln\sum_{n=n_{0}}^{n_{m}}\left(\beta\alpha\right)^{n}/n!$
, which reduces to $\mathcal{A}=\alpha\beta$ at large cutoff, as
one expects.

\subsection{Time evolution}

Just as with pure states, the evolution of a mixed state density matrix
can be treated in three ways: using differential identities, discrete
identities, or with a hybrid approach, combining linear differential
terms and nonlinear discrete matrix evolution for CPS mappings. The
hybrid approach is useful in some cases, since it removes the large
sampling error that can happen if time-evolution is treated using
stochastic equations derived purely from a Fokker-Planck equation. 

Consider a Liouvillian that includes loss and decoherence, with both
linear and nonlinear terms, so that: 
\begin{equation}
\mathcal{L}=\mathcal{L}^{(0)}+\mathcal{L}^{(n)}\,,
\end{equation}
where $\mathcal{L}^{(0)}$ describes the linear evolution and $\mathcal{L}^{(n)}$
the nonlinear part. Usually one must calculate time-evolution using
a master equation when there is coupling to reservoirs and dissipation,
so that:
\[
\mathcal{L}\left[\hat{\rho}\right]=\frac{-i}{\hbar}\left[\hat{H},\hat{\rho}\right]+\sum_{j}\gamma_{j}(2A_{j}\hat{\rho}A_{j}^{\dagger}-A_{j}^{\dagger}A_{j}\hat{\rho}-\hat{\rho}A_{j}^{\dagger}A_{j})\,\,,
\]
 where $\hat{H}$ is the system Hamiltonian, $\gamma_{j}$ are a set
of decay rates, and $\hat{A}_{j}$ are the corresponding operators
that describe coupling to reservoirs that cause decoherence. In the
differential approach, time-evolution is then obtained by transforming
the operators acting on $\hat{\lambda}$ or $\hat{\Lambda}$ to differential
operators, integrating by parts if this is possible, then solving
the resulting equations.

In the hybrid evolution approach, the base coherent state amplitudes
$\alpha,\beta$ evolve according to linear differential identities,
while the phase indices$\text{q,q'}$ evolve according to nonlinear
discrete identities. This has an advantage compared to the usual interaction
picture, in that any type of linear evolution is easily treated in
this fashion, including losses described by a master equation. 

In greater detail, one expands using both the discrete indices and
the base amplitude, so that using Einstein summation over repeated
$\bm{q}$ indices:

\begin{equation}
\hat{\rho}=\int p_{\overrightarrow{\bm{q}}}\left(\overrightarrow{\bm{\alpha}}\right)\hat{\lambda}_{\overrightarrow{\bm{q}}}\left(\overrightarrow{\bm{\alpha}}\right)d\mu\,,\label{eq:Projected-P-definition-1-1}
\end{equation}

Under time-evolution,
\begin{equation}
\frac{\partial}{\partial t}\hat{\rho}=\int p_{\overrightarrow{\bm{q}}}\left(\overrightarrow{\bm{\alpha}}\right)\left\{ \mathcal{L}^{(0)}\left[\hat{\lambda}_{\overrightarrow{\bm{q}}}\right]+\mathcal{L}^{(n)}\,\left[\hat{\lambda}_{\overrightarrow{\bm{q}}}\right]\right\} d\mu\,\,.
\end{equation}

Next, suppose that one uses identities so that $\mathcal{L}^{(0)}\rightarrow\mathcal{D}\left(\overrightarrow{\bm{\alpha}}\right)$,
$\hat{H}^{(n)}\rightarrow\bm{\mathcal{H}}^{(n)}$. Provided partial
integration is possible with vanishing boundary terms, the resulting
time-evolution equation is:

\begin{equation}
\dot{p}_{\overrightarrow{\bm{q}}}\left(\overrightarrow{\bm{\alpha}}\right)=\left[\mathcal{D}'\left(\overrightarrow{\bm{\alpha}}\right)\delta_{\overrightarrow{\bm{q}}\overrightarrow{\bm{q}}'}+\mathcal{L}_{\overrightarrow{\bm{q}}\overrightarrow{\bm{q}}'}^{(n)}\right]p{}_{\overrightarrow{\bm{q}}'}\left(\overrightarrow{\bm{\alpha}}\right)\,,
\end{equation}
where the partially integrated operator $\mathcal{D}'\left(\overrightarrow{\bm{\alpha}}\right)$
reverses the order of differentiation and changes the sign of each
differential term. 

\section{Examples}

There are many ways to use these techniques, depending on the projections
and symmetries involved. It is impossible to cover all the relevant
applications in one paper. The details in each case are subtly different,
and this needs to be taken into account in individual problems. In
this section, three simple illustrative examples are covered, based
on recent experiments in atom optics and photonics. The examples use
the CPS basis, but this is only one type of projection. Others are
also useful, and can be treated with the same general approach.

\subsection{Anharmonic oscillator coherent dynamics}

To understand how to use these techniques, consider the dynamical
evolution of a quantum anharmonic oscillator, which has both recurrences
and macroscopic superpositions. This is highly nontrivial to treat
using other phase-space methods \cite{Deuar2006a,Deuar2006b}, owing
to its nonclassical behavior. It is exactly soluble in the lossless
single-mode case \cite{Milburn:1986,drummond2014quantum}, and revivals
have been experimentally investigated \cite{Greiner:2002}. As such,
it is a useful test case, and the present techniques can be used to
extend these results to include real-world decoherence effects.

The Hamiltonian in the single-mode case is: 
\begin{equation}
\hat{H}/\hbar=\omega\hat{n}+\frac{1}{2}\kappa\hat{n}^{2}\,.\label{eq:anharmonicH}
\end{equation}
In the graphs below, the evolution of a projected coherent state $\left|\alpha\right\rangle _{\mathcal{Q}}$
is directly computed using a circular basis of CPS states with the
same radius, $r=\left|\alpha_{0}\right|$. Time-evolution results
are obtained by exponentiating the Hamiltonian matrix over a small
interval $\Delta t$ to give 
\begin{equation}
\bm{\psi}(t+\Delta t)=\exp\left(-i\bm{\mathcal{H}}^{A}\Delta t\right)\bm{\psi}(t)\,,
\end{equation}
where the Hamiltonian matrix $\bm{\mathcal{H}}^{A}$ is

\[
\bm{\mathcal{H}}^{A}=\omega\bm{\mathcal{O}}^{[1]}+\frac{1}{2}\kappa\bm{\mathcal{O}}^{[2]}\,.
\]
Apart from round-off errors, this is an exact procedure.

The analytical solution has an especially simple form for the average
coherent amplitude with no projection, 
\begin{equation}
\left\langle \hat{a}\right\rangle _{\infty}=\alpha\exp\left[\left|\alpha\right|^{2}\left(e^{-i\kappa t}-1\right)-i\left(\omega+\kappa/2\right)t\right]
\end{equation}

\begin{figure}
\centering{}\includegraphics[clip,width=0.9\columnwidth]{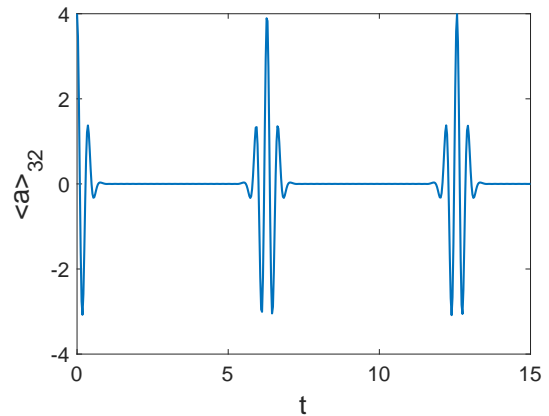}\caption{Anharmonic evolution of a coherent phase state mean amplitude, $\left\langle \hat{a}\right\rangle $,
for an initial state with $\alpha_{0}=4$ and $d=32$. The graph shows
the CPS numerical solution, which is indistinguishable from an analytic
calculation with no cutoff.\label{fig:Anharmonic-amplitude}}
\end{figure}

The quantity plotted is the average coherent amplitude $\left\langle \hat{a}\right\rangle _{d}$
for $\alpha=4$ , with parameters $\chi=1$ and $\omega=0.5$ and
a projection of $n<d$. For these parameters, an exact revival occurs
at $t=2n\pi$, for $n=1,2,\ldots$. Numerical results are shown in
Fig (\ref{fig:Anharmonic-amplitude}), using a cutoff of at $d=2\alpha^{2}=32$
and $500$ time-steps\@.

\begin{figure}
\centering{}\includegraphics[clip,width=0.9\columnwidth]{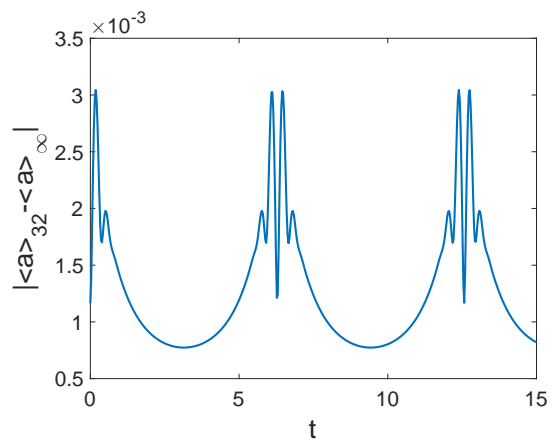}\caption{Change in the mean amplitude, $\left\langle \hat{a}\right\rangle $,
for anharmonic evolution of initial coherent state using CPS states
with $\alpha_{0}=4$ and $d=32$, showing the effects of the CPS number
cutoff. \label{fig:Anharmonic-error-1}}
\end{figure}

A graph of the difference between the projected and analytic result
is shown in Fig (\ref{fig:Anharmonic-error-1}), showing that the
maximum change in amplitude due to the number cut-off is $3\times10^{-3}$.
This shows that these techniques can be readily applied to a model
of nonlinear quantum dynamics.

\begin{figure}
\centering{}\includegraphics[clip,width=0.9\columnwidth]{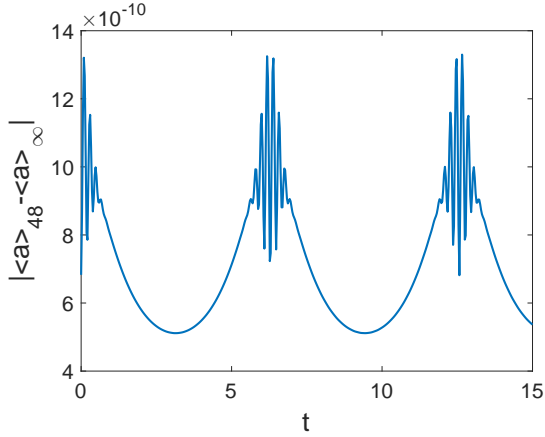}\caption{Effects of the number cut-off, for anharmonic evolution of initial
coherent state using CPS states with $\alpha_{0}=4$ and $d=48$,
showing reduced discrepancies as small as $10^{-9}$. \label{fig:Anharmonic-error-2}}
\end{figure}

The approach can be easily extended to larger cutoff values. Fig (\ref{fig:Anharmonic-error-2})
shows the effect of changing both cutoff and using the hybrid picture,
so that the linear evolution takes place using an interaction picture,
in which the unitary matrix $U$ includes a nonlinear Hamiltonian
$H_{n}=\kappa(\hat{n}-n_{m}/2)^{2}/2$, together with a renormalized
linear detuning of $\tilde{\omega}=\omega+\kappa n_{m}/2$ . As expected,
using the hybrid picture makes little or no difference: both forms
are equally exact. 

In this calculation, there is also a $50\%$ larger cutoff of $d=3\alpha^{2}=48$.
The maximum difference is now reduced to $1.4\times10^{-9}$. This
difference can either be viewed as the difference between an ideal
and practical situation, due to a physical limit to the photon number
from saturation effects. Alternatively the difference can be viewed
as a numerical artifact of using a number cutoff basis as an approximate
representation of a coherent state. 

In all cases, computed normalization errors were negligible, at most
of order $\sim10^{-12}$, caused by roundoff in $64$ bit IEEE arithmetic
and error propagation. This demonstrates that a CPS basis is able
to generate accurate results with either type of identity, and with
negligible errors at large cutoff. The advantage is that CPS methods
are well suited to complex, multi-mode problems, since the inter-mode
coupling between two spatially localized modes is typically linear,
with simple coherent identities. These problems will be treated elsewhere.

\subsection{Generation of cat-states and decoherence through phase noise}

The anharmonic Hamiltonian Eq. (\ref{eq:anharmonicH}) has been analyzed
by Yurke and Stoler \cite{yurke1986} for the purpose of generating
Schr\"odinger cat-states \cite{schrodinger1935} . They denoted
the nonlinearity by the symbol $\Omega=\frac{\kappa}{2}$ and solved
in the interaction picture without decoherence. After a time $t=t_{c}=\frac{\pi}{2\Omega}$,
a system prepared in a coherent state $|\alpha\rangle$ , where$\alpha$
is real, was shown to evolve to the cat-state 
\begin{equation}
|\mathrm{cat}\rangle=\frac{1}{\sqrt{2}}\bigl(e^{-i\pi/4}|\alpha\rangle+e^{i\pi/4}|-\alpha\rangle\bigr).\label{eq:cat}
\end{equation}
This macroscopic superposition state is of much interest and for
small $\alpha$ has been generated experimentally using different
techniques. This is simply the superposition of two coherent states
with a relative phase of $\phi=\pi$, and is therefore described exactly
in the CPS basis. 

However, the time-evolution due to the anharmonic Hamiltonian (\ref{eq:anharmonicH})
needs to be modified in many real-world situations. For example, the
present techniques can be used to treat arbitrary nonlinearities of
the form $\hat{n}^{p}$, by employing the general identities of Eq
(\ref{eq:general-identities}). 

In ultra-cold atomic physics and photonics, a strong limitation on
such experiments is decoherence originating in coupling to external
reservoirs. The most common types are losses and/or fluctuating external
potentials, which give rise to phase noise. These can be modeled using
master equation methods in the limit of broad-band or Markovian reservoirs,
or by using the more realistic case of non-Markovian phase-noise.
Either - or a combination of both - can be readily solved using projected
coherent states.

Since the problem is now dissipative, it is commonly treated using
a theory of the full density matrix, $\hat{\rho}.$ In the Markovian
limit of broad-band phase noise, the resulting master equation has
the general Lindblad form:

\begin{align}
{\frac{d\hat{\rho}}{dt}} & =\frac{1}{i\hbar}\left[\hat{H}_{sys},\hat{\rho}\right]+\gamma_{p}(2\hat{n}\hat{\rho}\hat{n}-\hat{n}^{2}\hat{\rho}-\rho\hat{\rho}\hat{n})\,\nonumber \\
 & +\gamma_{a}(2\hat{a}\hat{\rho}\hat{a}^{\dagger}-\hat{a}^{\dagger}\hat{a}\hat{\rho}-\hat{\rho}\hat{a}^{\dagger}\hat{a})\,\,.
\end{align}
The dissipation terms here describe phase decoherence, $\gamma_{p}$
and absorptive loss, $\gamma_{a}$, and one can divide the time-evolution
into linear and nonlinear parts: 
\begin{equation}
{\frac{d\hat{\rho}}{dt}}=\mathcal{L}^{(n)}\left[\hat{\rho}\right]+\mathcal{L}^{(0)}\left[\hat{\rho}\right]\,.\label{eq:rho}
\end{equation}
where $\mathcal{L}^{(0)}\left[\hat{\rho}\right]=-i\left[\frac{1}{2}\kappa\hat{n}^{2},\rho\right]$,
and the linear Liouvillian is
\begin{equation}
\mathcal{L}^{(0)}\left[\hat{\rho}\right]=-i\omega\left[\hat{n},\hat{\rho}\right]+\sum_{j}\gamma_{j}(2A_{j}\hat{\rho}A_{j}^{\dagger}-A_{j}^{\dagger}A_{j}\hat{\rho}-\hat{\rho}A_{j}^{\dagger}A_{j})\,\,.\label{eq:linmaster}
\end{equation}

The effects of absorptive loss in the Markovian limit have been treated
using the Q-function approach \cite{MilHolmesPhysRevLett.56.2237}.
In this paper we will use our techniques to predict the effect of
decoherence in the important case of general, non-Markovian phase
noise as opposed to absorptive losses; although both can be included
if necessary. With non-Markovian phase-noise, the master equation
must be replaced by an average over coherent evolution equations,
each with a different fluctuating linear term $\omega(t)$, and functional
probability $f[\omega]$.

Yurke and Stoler proposed as a signature for the cat-state the observation
of interference fringes in the probability distribution $P(p)$ where
$p$ is the result of a measurement of the quadrature phase amplitude
$\hat{p}=\frac{1}{i\sqrt{2}}[\hat{a}-\hat{a}^{\dagger}]$. The probability
distribution $P(x)$ for measurement of the orthogonal quadrature
$\hat{x}=\frac{1}{\sqrt{2}}[\hat{a}+\hat{a}^{\dagger}]$ is (for $\alpha$
real) a two-peaked Gaussian, with one hill located at $x=-\sqrt{2}\alpha$
and the other located at $\sqrt{2}\alpha$. The variances $(\Delta x)_{-}^{2}$
and $(\Delta x)_{+}^{2}$ associated with each hill are given by the
variance of a coherent states $|-\alpha\rangle$ and $|\alpha\rangle$:
$(\Delta x)_{\pm}^{2}=\frac{1}{2}$. As $\alpha\rightarrow\infty$,
the two hills become macroscopically distinguishable, and there is
a simplistic analogy of the `` cat'' being measured as ``alive''
or ``dead''. The two-peaked distribution for $P(x)$ could also
be generated by the classical mixture of the two coherent states 
\begin{equation}
\rho_{mix}=\frac{1}{2}[|\alpha\rangle\langle\alpha|+|-\alpha\rangle\langle-\alpha|]\label{eq:clasmix}
\end{equation}
Yurke and Stoler point out that this classical mixture (\ref{eq:clasmix})
however would not give the interference fringes in $P(p)$. The fringes
thus distinguish the mixture (\ref{eq:clasmix}) from the superposition
(\ref{eq:cat}). For $\alpha$ real
\begin{equation}
\langle p|\alpha\rangle=\frac{\exp(-i\sqrt{2}p\alpha+\left[\alpha^{2}-p^{2}-\left\Vert \alpha\right\Vert ^{2}\right]/2\}}{\pi^{1/4}}\label{eq:palphaproj}
\end{equation}
and hence the distribution for the classical mixture of the coherent
states is $P(p)=\frac{1}{\sqrt{\pi}}\exp{(-p^{2})}$, whereas for
the superposition it is:
\begin{equation}
P(p)=\frac{e^{-p^{2}}}{\sqrt{\pi}}[1-\sin(2\sqrt{2}\alpha p)]\label{eq:probfringe}
\end{equation}
To evaluate the fringe pattern accounting for the effect of arbitrary
phase noise, $\omega(t)$, we note that using the hybrid picture of
Eq (\ref{eq:hybrid}), the time-evolution in the CPS basis can be
factorized into linear evolution of the reference amplitude $\alpha(t)$,
together with a unitary transformation on the discrete phase indices
$q$. The reference amplitude only depends on the time integral of
the frequency or accumulated phase 
\begin{equation}
\theta(t)=\int_{0}^{t}\omega(t')dt'.
\end{equation}

The fringe pattern at the critical time $t=t_{c}$ is then given by
the phase distribution over the accumulated phase $\theta_{c}=\theta\left(t_{c}\right)$,
taking into account that at this time only two CPS states contribute:
\begin{align}
P(p) & =\int f(\theta_{c})\left|\sum_{q}\psi_{q}\langle p|\left\Vert \alpha e^{i\left(q\phi-\theta_{c}\right)}\right\rangle _{\mathcal{Q}}\right|^{2}d\theta_{c}\nonumber \\
 & =\frac{1}{2}\int f(\theta_{c})\left|\sum_{q}\psi_{q}\langle p|\left(|\alpha_{c}\rangle+i|-\alpha_{c}\rangle\right)\right|^{2}d\theta_{c}\,,
\end{align}
where $\alpha_{c}=\alpha e^{-i\theta_{c}}$. 

\begin{figure}
\centering{}\includegraphics[clip,width=0.9\columnwidth]{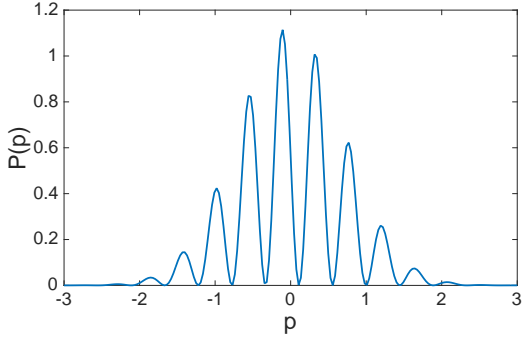}\caption{Interference fringes for the anharmonic oscillator at $t=t_{c}$,
$\alpha=5$. \label{fig:Anharmonic-cat}}
\end{figure}

\begin{figure}
\centering{}\includegraphics[clip,width=0.9\columnwidth]{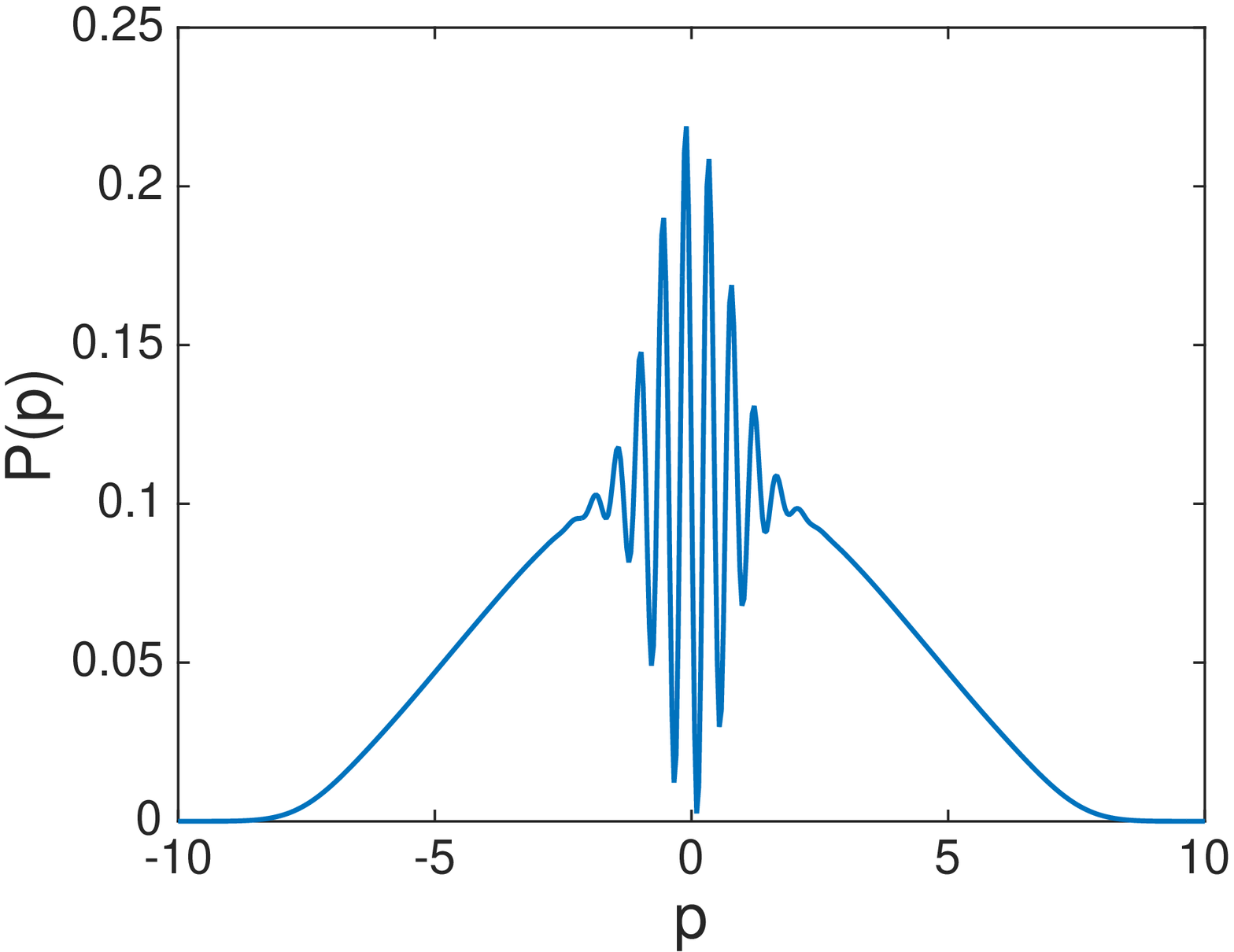}\caption{Interference fringes for the anharmonic oscillator at $t=t_{c}$,
$\alpha=5$, including phase noise with accumulated phase standard
deviation of $\sigma=0.5$. \label{fig:Anharmonic-cat0.5}}
\end{figure}

\begin{figure}
\centering{}\includegraphics[clip,width=0.9\columnwidth]{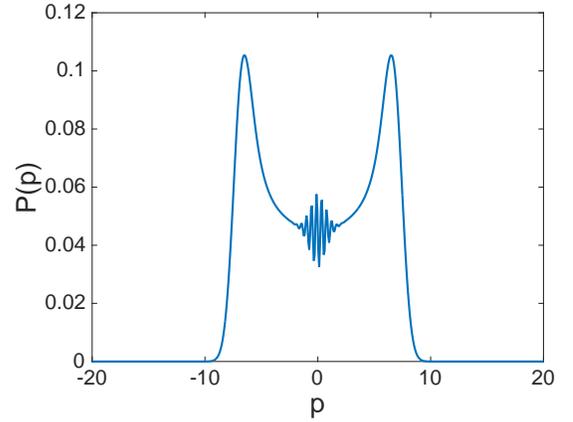}\caption{Interference fringes for the anharmonic oscillator at $t=t_{c}$,
$\alpha=5$, including phase noise with accumulated phase standard
deviation of $\sigma=2.0$. \label{fig:Anharmonic-cat2}}
\end{figure}

Results obtained using different levels of phase noise are plotted
in Figs (\ref{fig:Anharmonic-cat}), (\ref{fig:Anharmonic-cat0.5})
and (\ref{fig:Anharmonic-cat2}), with an initial coherent amplitude
of $\alpha=5$, and phase standard deviations of $\sigma=0$, $\sigma=0.5$
and $\sigma=2$ respectively. These results were obtained with a set
of $500$ points in momentum, and a Gaussian ensemble of $10^{5}$
random phases. They show the destruction and broadening of the fringe
pattern due to phase decoherence as the level of phase noise increases. 

\subsection{Application to boson sampling and photonic networks}

Recent experiments on photonic networks make use of single-photon
inputs into a linear, multimode photonic device. A typical experiment
excites $N$ input modes of an $M$ mode linear device. This is an
$M$ mode beamsplitter, with arbitrary phase-shifts and arbitrary
beam-splitting operations. The overall effect is to unitarily transform
the $M$ input modes into $M$ output modes. In addition, there are
some losses, although these can be modeled as undetected channels. 

The output correlations of up to $N-$th order are measured. Well-known
results in computational complexity theory indicate that the $N$-th
order correlations, proportional to the square of a matrix permanent,
are exponentially hard to compute exactly \cite{Valiant1979}. It
is conjectured that the resulting random samples of photon counts
are exponentially complex to generate on a classical computer \cite{Aaronson2011}.
This is called the 'Boson Sampling' problem. Other metrology applications
of these devices are also being investigated, using interferometric
schemes \cite{Motes2015_PRL114}.

Phase-space techniques are able to calculate the output correlations
very efficiently, although not exactly, owing to the \#P hardness
of permanents. To achieve this, one must have a representation of
the relevant input states. The full contour integral based existence
theorem for a arbitrary number state can do this. However, the inputs
in many current experiments are binary, with at most a single photon
input, so that $n_{j}=0,1$. Such experiments are well suited to the
projected P-representation. 

Since the initial photon number is bounded, it is useful to consider
a projective construction, valid in the limit of $r\rightarrow0$.
In this limit the ordinary complex P-representation and the projected
complex P-representation are identical, and projected methods allow
us to define an alternative existence theorem without changing the
basis set. For the qubit case with $n_{i}=0,1$, in each mode and
input dimension $d=2$, one sees from Eq (\ref{eq:phase-choice})
that  $\phi_{i}=\{0,\pi\}$, which implies that $\alpha_{i}=\pm r$. 

As explained above, with this expansion, a complex qudit P-function
$P_{Q}$ always exists.  In using the small radius limit, all the
known identities for the generalized P-representation are all still
valid, provided the $r\rightarrow0$ limit is taken after the calculation.
Thus, one can use the standard result that after transmission through
a linear optical system with phase-shifts, beam-splitters and losses,
the only effect on the representation is that the output coherent
amplitudes are multiplied by the relevant linear transmission matrix.
In calculating $N-th$ order correlations of an $N-$photon input,
all the factors proportional to the radius $r$ simply cancel. The
advantage of the present approach is that it gives unbiased results
for the observable modulus squared of the permanent, and can be readily
extended to include more complex inputs, outputs and nonlinear effects.

Complete details and numerical results for boson sampling are given
elsewhere \cite{Opanchuk2016-quantum}, showing the qubit basis is
particularly useful for treating phase noise effects in boson sampling
interferometry in the challenging, large $N$ limit with $N$ up to
$100$. Because of the exponential hardness of permanent calculations,
this is not viable using conventional permanent algorithms above $N=50$,
even on the largest current supercomputers.

\section{Conclusion}

Results are derived for coherent states in projected spaces. These
include the definition of a finite set of coherent phase states, called
coherent phase states, which are linearly independent and provide
a unique basis for expansions of arbitrary projected states. 

These methods allow a new type of generalized P-representation to
be introduced, applicable to a projected Hilbert space. Both existence
theorems and operator identities applicable to this projected representation
are derived. Results are obtained that allow changes of basis. 

Calculations are demonstrated with an anharmonic oscillator example,
including the effects of phase decoherence on the formation of a Schr\"odinger
cat. As another example,  the technique can be readily applied to
exponentially complex photonic network experiments.
\begin{acknowledgments}
This research was supported in part by the National Science Foundation
under Grant No. NSF PHY-1125915, and by the Australian Research Council.
\end{acknowledgments}

\section*{}

\bibliographystyle{apsrev4-1}
\bibliography{RMP_Draft_references,BellQsims,BosonRefs,ProjectP_references}

\end{document}